\documentclass[a4paper,12pt,dvips]{article}
\usepackage{amssymb}
\usepackage{amsmath}
\usepackage{graphics}
\usepackage{color}

\input{psfig}
\newcommand{\be}{\begin{equation}}
\newcommand{\ee}{\end{equation}}

\newcommand{\beqa}{\begin{eqnarray}}
\newcommand{\eeqa}{\end{eqnarray}}

\renewcommand{\bar}{\overline}
\title{The inverse problem for the Gross--Pitaevskii equation\footnote{~To be published in ``Chaos'' in 2010.}}
\author{Boris A. Malomed$^{1}$ and Yury A.~Stepanyants$^{2}$\footnote{~Corresponding author, e-mail: yuas50@gmail.com}}
\date{\small $^{1}$Department of Interdisciplinary Studies, School of Electrical
Engineering, Faculty of Engineering, Tel Aviv University, Tel Aviv, Israel; \\
$^2$Department of Mathematics and Computing, Faculty of Science, University of
Southern Queensland, Toowoomba, Australia.}

\begin{document}

\maketitle

\vspace{1cm}

\begin{abstract}
%\vspace*{20mm}
Two different methods are proposed for the generation of wide classes of
exact solutions to the stationary Gross--Pitaevskii equation (GPE). The
first method, suggested by the work by Kondrat'ev and Miller (1966), applies
to one-dimensional (1D) GPE. It is based on the similarity between the GPE
and the integrable Gardner equation, all solutions of the latter equation
(both stationary and nonstationary ones) generating exact solutions to the
GPE, with the potential function proportional to the corresponding
solutions. The second method is based on the ``inverse problem" for the GPE,
i.e. construction of a potential function which provides a desirable
solution to the equation. Systematic results are presented for 1D and 2D
cases. Both methods are illustrated by a variety of localized solutions,
including solitary vortices, for both attractive and repulsive nonlinearity
in the GPE. The stability of the 1D solutions is tested by direct
simulations of the time-dependent GPE.
\end{abstract}

\clearpage

\section{Introduction}
\label{sect1}

The Gross--Pitaevskii equation (GPE) provides for an exceptionally accurate
description of the dynamics of Bose--Einstein condensate (BEC) \cite%
{PitString03}. In the general case, the GPE is far from integrability, which
was an incentive for the development of various methods for simulations of
this equation, including finite-difference \cite{finite}, split-step \cite%
{split-step}, and Crank--Nicolson \cite{Crank} algorithms. An efficient
technique for finding \emph{stationary solutions} to the GPE is based on
simulations of the evolution in the imaginary time \cite{Chiofalo00}. A
review of numerical methods for the GPE can be found in Ref. \cite%
{Minguzzi04}.

Aside from the numerical solutions, the understanding of results produced by
the GPE requires the knowledge of its solutions in an analytical form --
approximate or, if possible, exact. A powerful analytical method is provided
by the variational approximation \cite{VA}. Another approach which
simplifies the consideration reduces the three-dimensional (3D) GPE to an
effective 1D or 2D form, if the condensate is loaded, respectively, into a
cigar-shaped or pancake-shaped trapping potential \cite{Luca}. If the
condensate is trapped in a deep optical-lattice (OL) potential, the
continual GPE may be further reduced to its discrete version \cite{discrete}%
. In the case of the repulsive nonlinearity, the Thomas--Fermi approximation
is known to be very useful \cite{PitString03,TF}. The coupled-mode
approximation is adequate for the description of settings based on double-
and multi-well potentials \cite{DWP}. In the case when the GPE contains a
rapidly oscillating time dependence, one may apply the averaging
approximation \cite{Fatkhulla}. If terms which make the 1D GPE different
from the exactly integrable nonlinear Schr\"{o}dinger equation (NLSE) are
small, one may resort to perturbation theory based on the inverse-scattering
transform \cite{RMP,PTIST}. A number of other approximations have been
developed in the context of the GPE, as reviewed in Ref. \cite{review}.

Although exact solutions of the GPE are rare, they are useful in those cases
when they are available (see. e.g., Ref. \cite{Spain}). An example is a
family of exact stationary periodic solutions to the 1D GPE with a specially
devised periodic potential, written in terms of elliptic functions. This
analysis was performed for both cases of the repulsive \cite{Carr-rep} and
attractive \cite{Carr-attr} nonlinearity in the GPE, see also Ref. \cite%
{Seaman}. Exact solutions were also found for dark-soliton trains
representing the supersonic flow of the condensate \cite{Kamch}. Exact
localized solutions are known in the case when the nonlinearity coefficient
is represented by a delta-function of the spatial coordinate \cite{delta1}
and by a symmetric pair of delta-functions \cite{delta2}. In fact, the
latter configuration provides for an exact solution to the
spontaneous-symmetry-breaking problem. Upon a proper change of the notation,
the GPE may be interpreted as the NLSE for spatial optical beams in
nonlinear waveguides \cite{KA}. Accordingly, the exact solutions found in
terms of the GPE may also find applications to nonlinear optics.

The purpose of this work is to propose two methods for generating exact
solutions to the GPE, together with potentials which support them. The first
method is based on the idea proposed by Kondrat'ev and Miller \cite%
{KondrMiller} more than 40 years ago, namely, using known
solutions of nonlinear equations as potentials for other equations
(a somewhat similar method was later proposed for the analysis of
self-trapped states in nonlinear optics \cite{Snyder}). We apply
it by noting that the 1D time-independent GPE with the potential
term is equivalent to a stationary Gardner equation (GE), alias an
extended Korteweg--de Vries (KdV) equation, containing both
quadratic and cubic nonlinearities, if the solution is
proportional to the potential. Thus, one can obtain a solution to
the GPE, along with the necessary potential, from any solution of
the GE.

The second method is based on the consideration of an \textit{inverse problem%
}, aiming to construct an appropriate potential for a \emph{given}
wave-function ansatz representing an appropriate solution. The inverse
problem is relevant because it may be relatively easy to engineer the needed
trapping potential, using external magnetic and optical fields \cite%
{trapping}. Recent experiments have demonstrated that, using a rapidly
moving laser beam focused on the condensate, one can ``paint" practically
any desired time-average potential profile in 1D and 2D settings \cite%
{painting}. A similar approach was developed in a different physical
setting, with the objective to find models of nonlinear dynamical chains
admitting exact solutions for traveling discrete pulses in an analytical
form \cite{Flach}. By means of this approach, we produce a number of novel
1D and 2D stationary solutions. The stability of the 1D solutions obtained
by both above-mentioned methods is tested in direct simulations.

The paper is organized as follows. In Section \ref{sect2}, we elaborate the
Kondrat'ev--Miller method in the 1D case and perform the stability test of
the localized modes. We show that stationary solutions of the GPE may be
constructed using not only stationary, but also \textit{nonstationary}
solutions of the GE. In the latter case, the use of the formal temporal
variable in the GE makes it possible to obtain a wide family of stationary
solutions and supporting potentials which depend on a continuous parameter,
the relation between them being different from the proportionality. In the
same section, we present the inverse method in the 1D case. The stability of
these solutions is tested via direct simulations of the time-dependent GPE.
In Section \ref{sect3}, we construct exact solutions to the GPE in the 2D
case (the test of their stability will be reported elsewhere). In
particular, we construct anisotropic solutions, using the so-called lump
solitons of the Kadomtsev--Petviashvili equation (KP1) as the respective
ansatz, axisymmetric states with the Gaussian radial profile, and vortices
with topological charges $1$ and $2$. The paper is concluded by Section 4.

\section{Exact solutions of the one-dimensional Gross--Pitaevskii equation}

\label{sect2}

The scaled form of the 1D GPE for complex wave function $\varphi (\xi ,t)$
is well known \cite{PitString03}:

\begin{equation}
i\varphi _{t}+\mu \varphi =-\varphi _{\xi \xi }+u(\xi )\varphi +\sigma
|\varphi |^{2}\varphi ,  \label{dl1DGPE}
\end{equation}%
where $u(\xi )$ is the trapping potential, $\sigma =+1$ or $-1$
corresponding to the repulsive and attractive interactions between atoms.
The constant $\mu $ in Eq.~(\ref{dl1DGPE}) is the chemical potential, the
objective being to find stationary solutions corresponding to a given value
of $\mu $. If Eq. (\ref{dl1DGPE}) is derived from the underlying GPE for the
3D condensate, the temporal variable $t$ and spatial coordinate $\xi $,
together with function $\varphi $ and normalized potential $u(\xi )$, are
related to their counterparts measured in physical units, $T,X$ and $\Psi $,
as follows: $t\equiv T\left( \pi ^{2}\hslash /md^{2}\right) $, $\xi \equiv
\pi X/d$, and
\begin{equation}
\Psi (X,R,T)=\sqrt{\frac{\pi }{2\left\vert a_{s}\right\vert d^{2}}}\,\varphi
(\xi ,t)\exp {\left( -i\omega _{\perp }T-\frac{\omega _{\perp }m}{2\hbar }%
R^{2}\right) },  \label{Psi}
\end{equation}%
\begin{equation}
U(X) = \frac{1}{m}\left(\frac{\pi \hbar}{d}\right) ^{2}u(\xi ),  \label{U}
\end{equation}%
where $m$ is the atomic mass, $d$ is a longitudinal scale determined by the
axial potential, $a_{s}$ is the $s$-wave scattering length, while $\omega
_{\perp }$ and $R$ are the transverse trapping frequency and radial
coordinate. If $m$ is taken for $^{87}$Rb, and $d=1.5$ $\mathrm{\mu }$m,
then $t=1$ and $\xi =1$ correspond, in physical units, to $T\approx 0.3\ $ms
and $X\approx 0.5$ $\mathrm{\mu }$m, respectively. Finally, the number of
atoms in the condensate is%
\begin{equation}
N\equiv 2\pi \int\limits_{0}^{\infty }R\,dR\int\limits_{-\infty
}^{+\infty }\left\vert \Psi (X,R,T)\right\vert ^{2}dX \equiv
\frac{\pi a_{\perp }^{2}}{2\left\vert a_{s}\right\vert
d}\int\limits_{-\infty }^{+\infty }|\varphi (\xi
)|^{2}d\xi , \label{N}
\end{equation}%
where the transverse-trapping size is $a_{\perp }=\sqrt{\hbar
/m\omega _{\perp }}$.

\subsection{Construction of stationary solutions by the \textit{%
Kondrat'iev--Miller method}}
\label{ss2.A}

Looking for real stationary solutions to Eq.~(\ref{dl1DGPE}), we arrive at
equation
\begin{equation}
\varphi ^{\prime \prime }+\mu \varphi =u(\xi )\varphi +\sigma \varphi \,^{3},
\label{St1DGPE}
\end{equation}%
with the prime standing for $d/d\xi $. Particular exact solutions to Eq. (%
\ref{St1DGPE}) can be obtained by way of the approach developed in Ref. \cite%
{KondrMiller}. To this end, parallel to Eq. (\ref{St1DGPE}), one should
consider the stationary GE (see \cite{AblSegur,Apel07} and references
therein):
\begin{equation}
\phi ^{\prime \prime }-V\phi =-\alpha \phi \,^{2}+\sigma \phi \,^{3},
\label{Gardner}
\end{equation}%
where $V$ and $\alpha $ are constants. It is well known that exact solutions
to Eq. (\ref{Gardner}) can be found in terms of the Jacobi's elliptic
functions. We chose one of such solutions and denote it $\Phi (\xi )$. Then,
the quadratic term in Eq.~(\ref{Gardner}) is formally factorized, and the
first multiplier is replaced by $\Phi (\xi )$, i.e. $\phi ^{2}\equiv \phi
\cdot \phi \rightarrow \Phi (\xi )\phi $. By substituting this into Eq.~(\ref%
{Gardner}), one obtains:
\begin{equation}
\phi ^{\prime \prime }-V\phi =-\alpha \Phi (\xi )\phi +\sigma \phi \,^{3}.
\label{GGPeq}
\end{equation}%
Obviously, this equation is immediately satisfied with $\phi =\Phi (\xi )$.
From here it follows that there is the one-to-one correspondence between
Eqs.~(\ref{St1DGPE}) and (\ref{GGPeq}):
\begin{equation}
\varphi \longleftrightarrow \phi, \quad \mu \longleftrightarrow -V, \quad
u(\xi )\longleftrightarrow -\alpha \Phi (\xi ).
\end{equation}
Thus, Eq.~(\ref{St1DGPE}) gives rise to the exact solution, $\varphi (\xi
)=\Phi (\xi )$, with chemical potential $-V$, for the external potential
\begin{equation}
u(\xi )=-\alpha \Phi (\xi ).  \label{u}
\end{equation}

\subsection{Stationary solutions in the case of the repulsive nonlinearity}
\label{ss2.B}

\textit{An illustrative example -- the ``fat soliton" of the Gardner
equation.} In the case of the repulsive atomic interactions, which
corresponds to $\sigma =+1$ in Eq. (\ref{dl1DGPE}), the approach outlined
above may be illustrated using a particular solution to Eq.~(\ref{Gardner})
in the form of so-called ``fat soliton" \cite{Apel07}):
\begin{equation}
\Phi (\xi )=\frac{\alpha \nu}{3}\left[ \tanh {\left( \xi /\Delta
+\theta \right) }-\tanh {\left( \xi /\Delta -\theta \right) }\right] ,
\label{FatSol}
\end{equation}%
\begin{equation}
\theta =(1/4)\ln \left[ (1+\nu )/(1-\nu )\right] ,~\Delta =3\sqrt{2}/(\alpha
\nu ),~V=(2/9)(\alpha \nu )^{2},  \label{Delta}
\end{equation}%
with free parameters $\alpha $ and $\nu $, the latter one taking values $%
0<\nu <1$. The front and rear slope of the fat soliton, $\Delta $, depends
monotonously on $\nu $, decreasing from infinity to $\Delta _{\min }=3\sqrt{2%
}/\alpha $ when $\nu $ varies from $0$ to 1. The width of the soliton, $L$,
i.e. the distance between its front and rear segments at the half-minimum
level, $\Phi (\xi )=\Phi _{\max }/2$, is
\begin{equation}
L=\frac{3\sqrt{2}}{\alpha \nu }\ln {\left[ 2+\left( 1-\nu ^{2}\right)
^{-1/2}+\sqrt{\left( 2+\left( 1-\nu ^{2}\right) ^{-1/2}\right) ^{2}-1}\,%
\right] }.  \label{FatSolWidth}
\end{equation}

\begin{figure}[h!]
\centerline{\psfig{figure=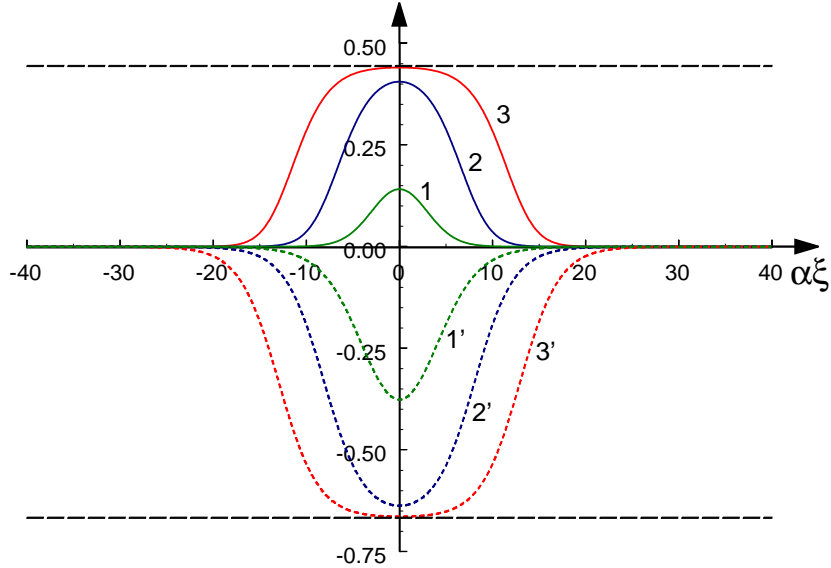,height=80mm}}
\caption{Normalized density $|\Phi (\protect\xi )/\protect%
\alpha |^{2}$ corresponding to solution (\protect\ref{FatSol}) (solid
lines), and the corresponding normalized potential $u(\protect\xi )/\protect%
\alpha ^{2}$ (broken lines), as functions of $\protect\alpha \protect\xi $.
Lines $1$ and $1^{\prime }$ pertain to $\protect\nu =0.9$; lines $2$ and $%
2^{\prime }$ -- to $\protect\nu =0.999$; lines $3$ and $3^{\prime }$ -- to $%
\protect\nu =0.99999$. The horizonal dashed lines show limit values of the
solution and potential function for the ``table-top soliton". }
\label{f01}
\end{figure}

At $\nu \rightarrow 0$, the fat soliton reduces to the bell-shaped KdV
soliton, whose width is given by $L_{\mathrm{KdV}}=3\sqrt{2}\ln {\left( 3+%
\sqrt{8}\right) }/(\alpha \nu )$. In the other limit, $\nu \rightarrow 1$,
it reduces to the ``table-top soliton" (a $\Pi $-shaped mode) with $L\approx
2\Delta \theta \approx -3\sqrt{2}\ln {\left[ {(1-\nu )/\nu }\right] }%
/(2\alpha \nu )$. The minimum width, $L_{\min }\approx 10.1/\alpha $, is
attained at $\nu \approx 0.892$. The local density corresponding to
normalized solution (\ref{FatSol}), $|\Phi /\alpha |^{2}$, along with the
corresponding normalized potential, $u/\alpha ^{2}$, in the stationary GPE
for which one has $\Phi (\xi )$ as the \emph{exact solution}, are shown in
Fig.~\ref{f01} for several values of free parameter $\nu $.

The dimensionless norm of exact solution (\ref{FatSol}), which is proportional to the
number of particles in the underlying BEC, according to Eq. (\ref{N}), is
$$%
N_{\mathrm{1D}}=\int\limits_{-\infty }^{+\infty }|\varphi (\xi )|^{2}\,d\xi =4\sqrt{%
2}\alpha \nu \left[\frac{1}{2\nu}\ln {\left(\frac{1 + \nu}{1 - \nu}\right) }-1\right].
$$

As follows from here, $N_{\mathrm{1D}}$
is proportional to $\alpha $, while $\nu $ may be treated as a free
parameter which determines the shape of the potential and the corresponding
solution. The chemical potential of the solution, $\mu =-V$, is also
determined by constants $\alpha $ and $\nu $, see Eq.~(\ref{Delta}).

Because wave function (\ref{FatSol}) has no zeros, it may represent the
\emph{ground state} in the corresponding potential, with chemical potential $%
\mu =-\left( 2/9\right) \left( \alpha \nu \right)^2 \ $[see Eq. (\ref{Delta}%
)]; whether there exist higher-order bound states with larger discrete
eigenvalues of $\mu $ within this nonlinear problem, remains an open
question. Although it is plausible that this solution is stable, it is
relevant to test its dynamics, under the action of perturbations, in direct
simulations of the time-dependent equation~(\ref{dl1DGPE}). This was done by
means of the Yunakovsky's method \cite{Yunakovsky} in a sufficiently large
domain with periodic boundary conditions (description of the method is
presented in the Appendix). Examples are shown in Fig.~\ref{f02} for cases
when the amplitude was initially reduced or increased by $10\%$ against the
stationary value.

\begin{figure}[h!]
\centerline{\psfig{figure=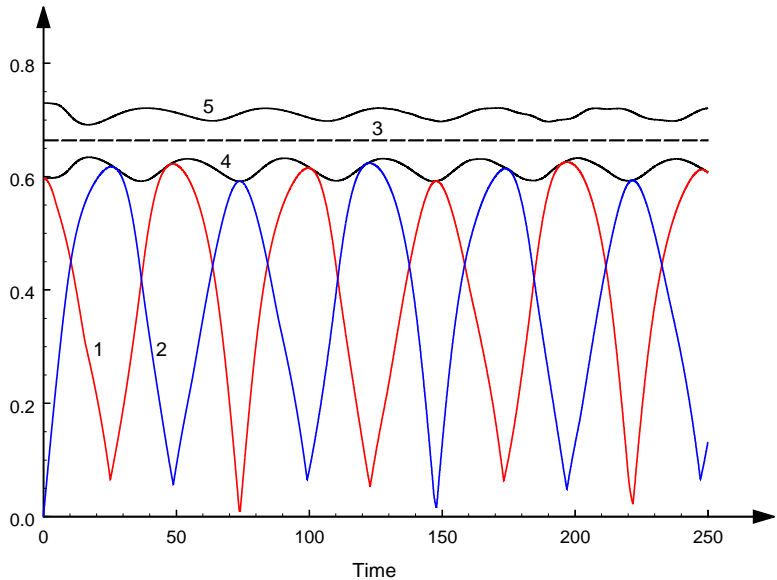,height=80mm}}
\caption{The time dependence of perturbed solution (\protect
\ref{FatSol}) in the model with the repulsive nonlinearity, when the initial
amplitude is $10\%$ smaller than needed for the stationary solution. Line 1
-- max$_{\protect\xi }|\mathrm{Re}\left\{ \protect\varphi (\protect\xi %
,t)\right\} |$, line 2 -- max$_{\protect\xi }|\mathrm{Im}\left\{ \protect%
\varphi (\protect\xi ,t)\right\} |$. Horizontal line 3 designates the
constant amplitude of stationary soliton (\protect\ref{FatSol}) with $%
\protect\nu =0.99999$. Lines 4 and 5 show the time dependence of max$_{%
\protect\xi }|\protect\varphi (\protect\xi ,t)|$ in the cases when the
soliton's amplitude was initially reduced (line 4) or increased (line 5) by $%
10\%$ against the stationary value.}
\label{f02}
\end{figure}

As seen from the figure, the amplitude of the so perturbed
solution varied in time within the same $10\%$, while its spatial shape was
preserved. It is also seen that the perturbation induced oscillations
between the real and imaginary parts of the solution, i.e. shifted its
chemical potential. In the course of the simulations, the norm of the
solution was preserved with relative accuracy $\sim 10^{-7}$. The latter
fact attests to the \emph{robustness} of the ground state: under the action
of this sufficiently strong perturbation, it features no loss through
emission of radiation.

Using the above physical estimates for BEC, it is easy to estimate
physical parameters of BEC\ states corresponding to the fat-soliton
solutions. For instance, the outermost configuration in Fig.
\ref{f01} represents to the potential well of the depth $\simeq 1$
recoil energy corresponding to $d=1.5$ $\mathrm{\mu }$m, and width
$L\simeq 30$, which is $\simeq 15$ $\mathrm{\mu } $m in physical
units (see above). These values are quite realistic for the
experiment \cite{PitString03,trapping,painting}. Further, taking an
experimentally relevant values of $a_{s}=5$ nm and $a_{\perp }=3$ $\mathrm{%
\mu }$m, Eq. (\ref{N}) yields the largest number of $^{87}$Rb
atoms which may form the fat soliton, $N\simeq 30,000$. This
estimate shows that the constructed soliton solution is quite
relevant to the experiment. Finally, the period of
large-amplitude oscillations of the perturbed stable state, shown in Fig. %
\ref{f02}, is $\sim 100$ ms.

\begin{figure}[h]
\centerline{\psfig{figure=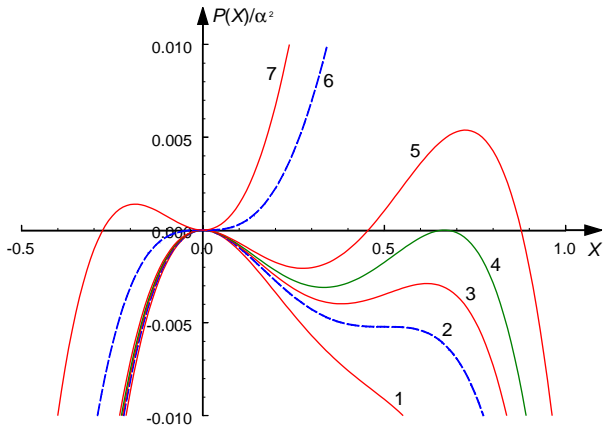,height=60mm}}
\caption{Normalized polynomial $P(X)/\protect\alpha ^{2}$
defined in Eq.~(\protect\ref{Potent1}) with $\protect\sigma =1$ at different
values of $W$: line $1$ -- for $W=9/32$, line $2$ -- for $W=1/4$, line $3$
-- for $W=17/72$, line $4$ -- for $W=2/9$, line $5$ -- for $W=1/5$, line $6$
-- for $W=0$, and line $7$ -- for $W=-2/9$.}
\label{f03}
\end{figure}

\textit{Other stationary solutions related to the Gardner equation}. To
consider various exact solutions to Eq. (\ref{Gardner}), one may write the
equation in the ``energy conservation" form, $(X^{\prime })^{2}/2+P(X)=E$,
where $X\equiv \phi /\alpha $, $E$ is a constant of integration, and $P(X)$
is the effective potential,
\begin{equation}
P_{+}(X)=-\left( \sigma /4\right) \alpha ^{2}X^{2}\left[ \left( X-2\sigma
/3\right) ^{2}+2\sigma W-4/9\right] ,  \label{Potent1}
\end{equation}%
with $W\equiv V/\alpha ^{2}$. It is shown in Fig.~\ref{f03} for $\sigma =1$.

For $W>1/4$, the polynomial has a single real maximum at $X=0$, hence
neither periodic nor solitary solutions are possible. At $W<1/4$, it has
three real extrema, at points $X=0,~(1/2)(1-\sqrt{1-4W}),~(1/2)(1+\sqrt{1-4W}%
)$. In that case, first appears a depression-type solitary solution, in the
form of a ``bubble" against the constant background value of the field, $%
\phi _{0}=(\alpha /2)\left( 1+\sqrt{1-4V/\alpha ^{2}}\right) $. The typical
potential profile corresponding to the ``bubble" is depicted by the line 3
in Fig.~\ref{f03}.

At $W=2/9$, the potential becomes symmetric, as shown in Fig.~\ref{f03} by
line 4. Solution (\ref{FatSol}) with $\nu =1$ corresponds exactly to this
case. For smaller values of $W$, when it varies from $2/9$ to $0$, the right
maximum of the potential function becomes taller then the left one, making
it possible to have bright solitons with the zero background. They
correspond to solution~(\ref{FatSol}) with $\nu <1$. The solution vanishes
at $W\rightarrow +0$, i.e. $\nu \rightarrow +0$. For $W<0$, the left maximum
of the potential shifts from the origin to the left, see Fig.~\ref{f03}). In
this case, solitons exist against the negative background, with $\phi
_{0}=(\alpha /2)\left( 1-\sqrt{1-4V/\alpha ^{2}}\right) $.

Periodic solutions of Eq.~(\ref{Gardner}) can be analyzed similarly. The
corresponding stationary solutions of the GPE can be readily obtained by
means of the method described above, in terms of elliptic functions, similar
to the periodic solutions reported in Ref. \cite{Seaman}.

\subsection{Stationary solutions in the case of the attractive nonlinearity}
\label{ss2.C}

\begin{figure}[th]
\begin{tabular}{cc}
\hspace*{-1cm}\psfig{figure=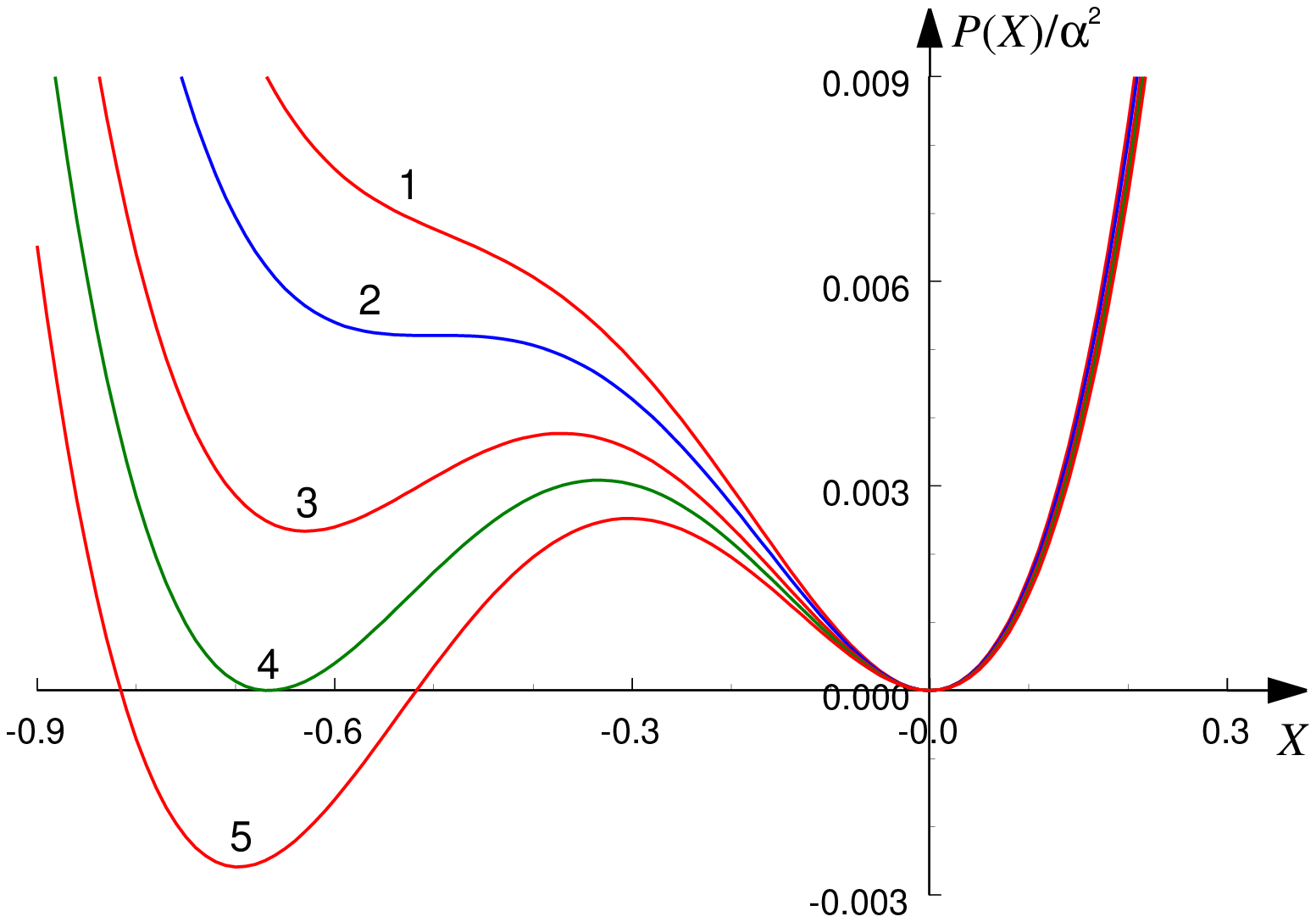,height=60mm} & \psfig{figure=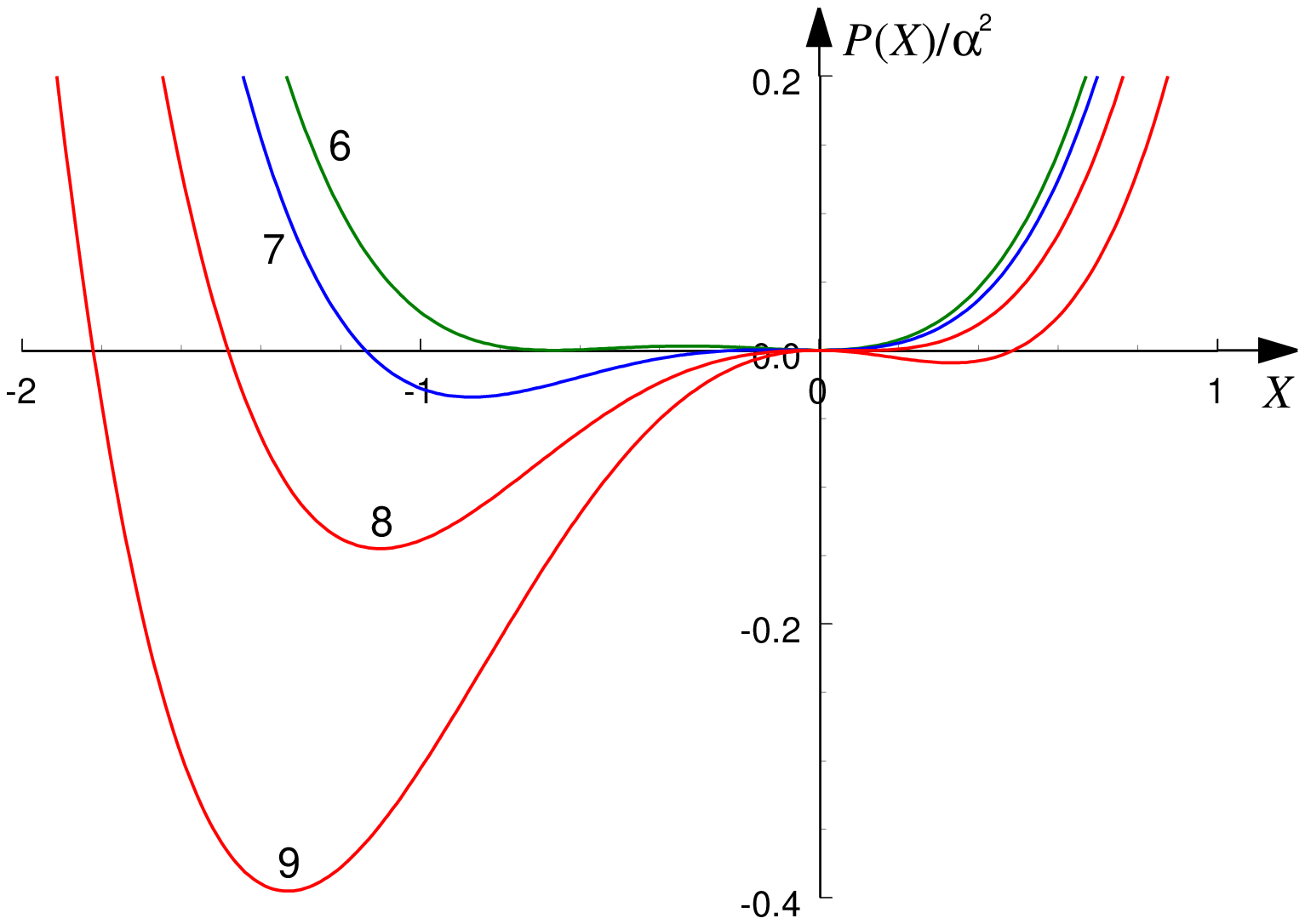,height=60mm}
\\
a) & b)%
\end{tabular}%
\caption{Normalized polynomial $P_{-}(X)/\protect\alpha ^{2}$
defined by Eq.~(\protect\ref{Potent1}) with $\protect\sigma =-1$ for
different values of $W$. Frame (a): line $1$ -- for $W=-0.26$, line $2$ --
for $W=-0.25$, line $3$ -- for $W=-0.23$, line $4$ -- for $W=-0.22$, and
line $5$ -- for $W=-0.21$. Frame (b) uses a different scale: line $6$ -- for
$W=-0.22$ [the same as line 4 in (a)], line $7$ -- for $W=-0.11$, line $8$
-- for $W=0.11$, and line $9$ -- for $W=0.44$.}
\label{f04}
\end{figure}

In the case of the attractive nonlinearity, i.e. $\sigma =-1$ in Eq.~(\ref%
{St1DGPE}), the potential function is different from that shown in Fig.~\ref%
{f03}. It is shown in Fig.~\ref{f04} in two different scales, as it is
impossible to display all details using a single scale. For $W\geq -1/4$,
three possible real extrema of this polynomial are located at points
$$%
X=0, \quad -\frac 12\left(1-\sqrt{1+4W}\right), \quad -\frac 12\left(1+\sqrt{1+4W}\right),
$$%
otherwise the polynomial
has a single real minimum at $X=0$, hence solitary solutions do not exist
for $W<-1/4$.

The first solitary-type solution emerges at $W=-1/4$. In this case, the
potential still has only one minimum at $X=0$, but the inflexion point
appears at $X=-1/2$ (the corresponding potential function is shown in Fig.~%
\ref{f04} by line 2). A particular solution corresponding to $W=-1/4$
represents the \textit{algebraic soliton} sitting on top of a pedestal
(constant-value background), as a solution to Eq.~(\ref{St1DGPE}) with free
parameter $\alpha $ and $V=-\mu =-\alpha ^{2}/4$:
\be%
\label{AlgSonPed}
\Phi (\xi ) = \frac{\alpha}{2}\left( \frac 43 \frac{1}{1+\alpha ^{2}\xi ^{2}/18} - 1\right).
\ee

\begin{figure}[h!]
\centerline{\psfig{figure=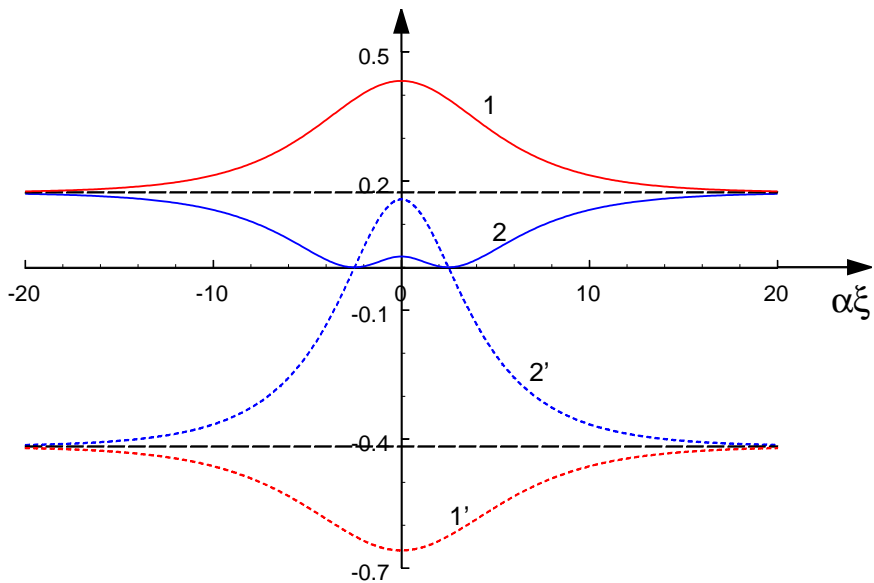,height=80mm}}
\caption{Solutions (\protect\ref{SolOnNegPed}) in terms of $%
(\Phi _{\pm }/\protect\alpha )^{2}$ (solid lines), and respective normalized
potentials $u/\protect\alpha ^{2}$ (broken lines) are shown as functions of
normalized coordinate $\protect\alpha \protect\xi $. Lines $1$, $1^{\prime }$
and $2$, $2^{\prime }$ pertain, respectively, to signs plus and minus in
Eq.~(\protect\ref{SolOnNegPed}). In both cases, $\protect\nu =-5/12$.}
\label{f05old}
\end{figure}

At $W>-1/4$, one more minimum appears in the potential profile (see, e.g.,
line 3 in Fig.~\ref{f04}). In this case, two families of solitons on a
pedestal are generated by Eq. (\ref{Gardner}):%
\begin{equation}
\Phi _{\pm }(\xi ) = \alpha \nu \left\{ 1-\frac{3(1+2\nu )}{1+3\nu \pm \sqrt{%
1+3\nu /2}\cosh {\left[ \alpha \xi \sqrt{-\nu (1+2\nu )}\,\right] }}\right\}
,  \label{SolOnNegPed}
\end{equation}%
where $\nu $ is a free parameter ranging between $-1/2$ and $0$, and $V=-\mu
=\alpha ^{2}\nu (1+\nu )$. The local densities corresponding to solution $%
\Phi _{+}(\xi )$ has the form of a double dark soliton (with two zeros),
whereas solution $\Phi _{-}(\xi )$ is shaped as a bump on top of the
pedestal. The local densities corresponding to solutions (\ref{SolOnNegPed}),
along with the respective potentials, for which these are
exact solutions to the GPE, are shown in Fig.~\ref{f05old} for $\nu =-5/12$.

When $W$ increases further and becomes equal to $-2/9$, potential function (%
\ref{Potent1}) becomes symmetric with respect to the vertical line $X=-1/3$
(see line 4 in Fig.~\ref{f04}), getting then asymmetric, with the left
minimum falling deeper than the right one when $W$ increases further (see,
e.g., line 5 in Fig.~\ref{f04}). For the particular case of $W=-2/9$,
solutions (\ref{SolOnNegPed}) reduce to
$$\Phi (\xi )=-\frac{\alpha}{3}\left[ 1 \pm \sqrt{2}\mathrm{sech}{\left(\frac{\alpha \xi}{\sqrt{3}}\right)}\right].
$$

\begin{figure}[t!]
\centerline{\psfig{figure=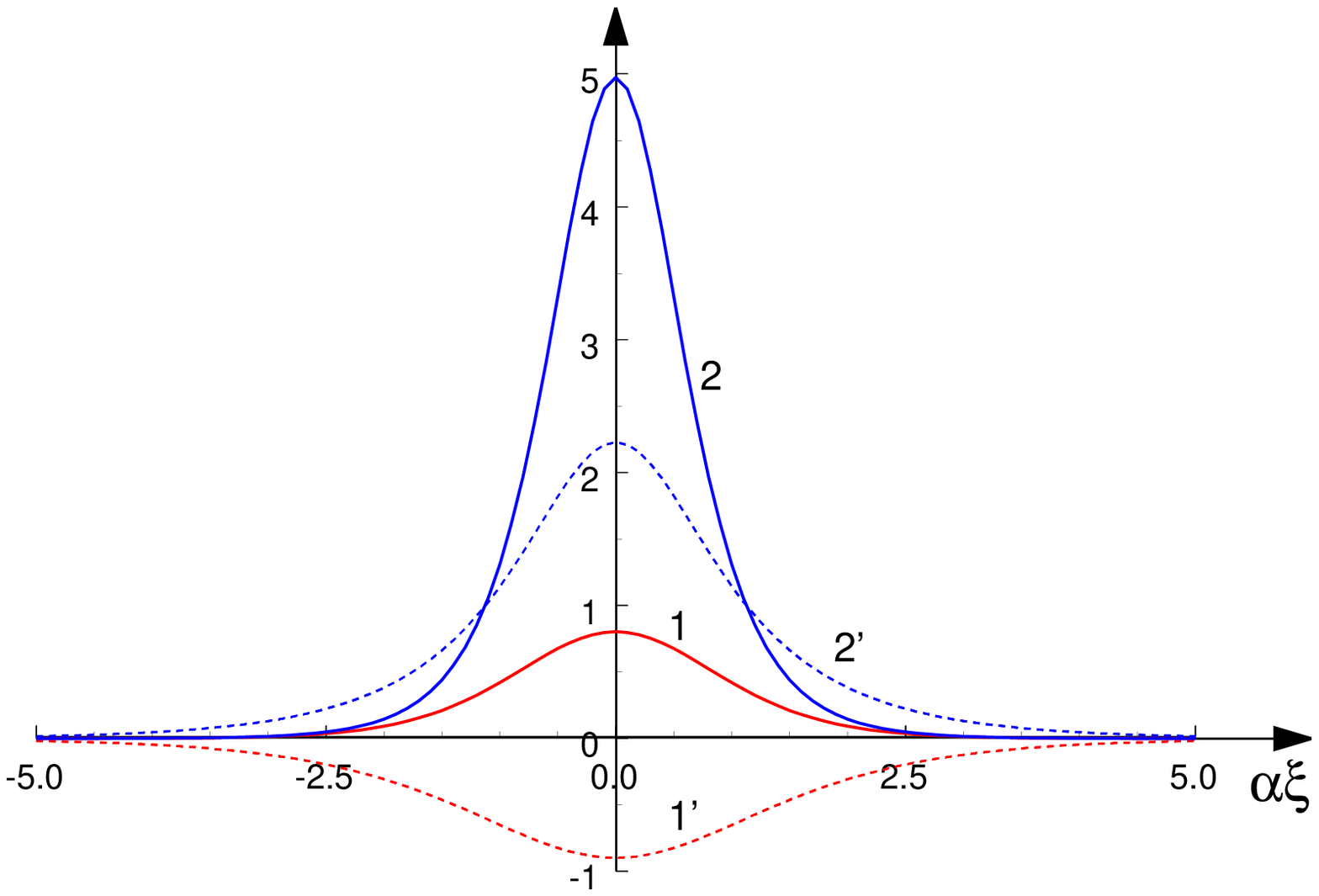,height=80mm}}
\caption{$(\Phi _{\pm }/\protect\alpha )^{2}$ for solutions (%
\protect\ref{ExpSol}) with $\protect\nu =1$ (solid lines), and the
corresponding normalized potentials, $u/\protect\alpha ^{2}$
(broken lines),
as functions of $\protect\alpha \protect\xi $. Lines $1$, $1^{\prime }$ and $%
2,$ $2^{\prime }$ pertain, respectively, to signs $+$ and $-$ in Eq.~(%
\protect\ref{ExpSol}).} \label{f05}
\end{figure}

For $W$ ranging between $-2/9$ and $-1/9$, the potential
function (\ref{Potent1}) becomes again asymmetric, as mentioned above, while
the corresponding solutions are still given by Eq.~(\ref{SolOnNegPed}) with $%
-1/3\leq \nu \leq 0$. They remain unstable, because of the nonzero
background.

At $W=-1/9$, the maximum and minimum of the potential merge at $X=0$.
Another inflexion point emerges in the potential profile in this case [see
line 7 in Fig.~\ref{f04}(b)]. The corresponding solution to Eq.~(\ref%
{Gardner}) is an algebraic soliton with $V=-\mu =0$ and zero background,
$$%
\Phi (\xi )=-\frac{4\alpha}{3}\frac{1}{1 + 2\alpha ^{2}\xi ^{2}/9}.
$$
However, as follows from Eq. (\ref{u}), this solution corresponds to
the maximum of the physical potential, hence its is unstable (which was
confirmed by direct simulations).

For $W>-1/9$, two families of \emph{exponentially localized} solitons are
generated by Eq. (\ref{Gardner}):
\begin{equation}
\Phi _{\pm }(\xi )=\frac{3\alpha \nu ^{2}}{1\pm \sqrt{1+9\nu ^{2}/2}\cosh {%
(\nu \alpha \xi )}},  \label{ExpSol}
\end{equation}%
where free parameter $\nu \geq 0$ determines the inverse width of the
soliton, and $V=-\mu =(\alpha \nu )^{2}$. These solutions and corresponding
potentials are shown in Fig.~\ref{f05} for $\nu =1$. As follows from Eq. (%
\ref{u}), solution $\Phi _{-}(\xi )$ is unstable, as it represents a soliton
sitting at the potential maximum. However, solution $\Phi _{+}(\xi )$ is
trapped in the minimum of the attractive potential, hence it may be stable.
At $\nu \rightarrow 0$, the latter solution smoothly vanishes, whereas the
unstable one reduces to the above-mentioned unstable algebraic soliton.

\begin{figure}[ht]
\vspace{-20mm}
\centerline{\psfig{figure=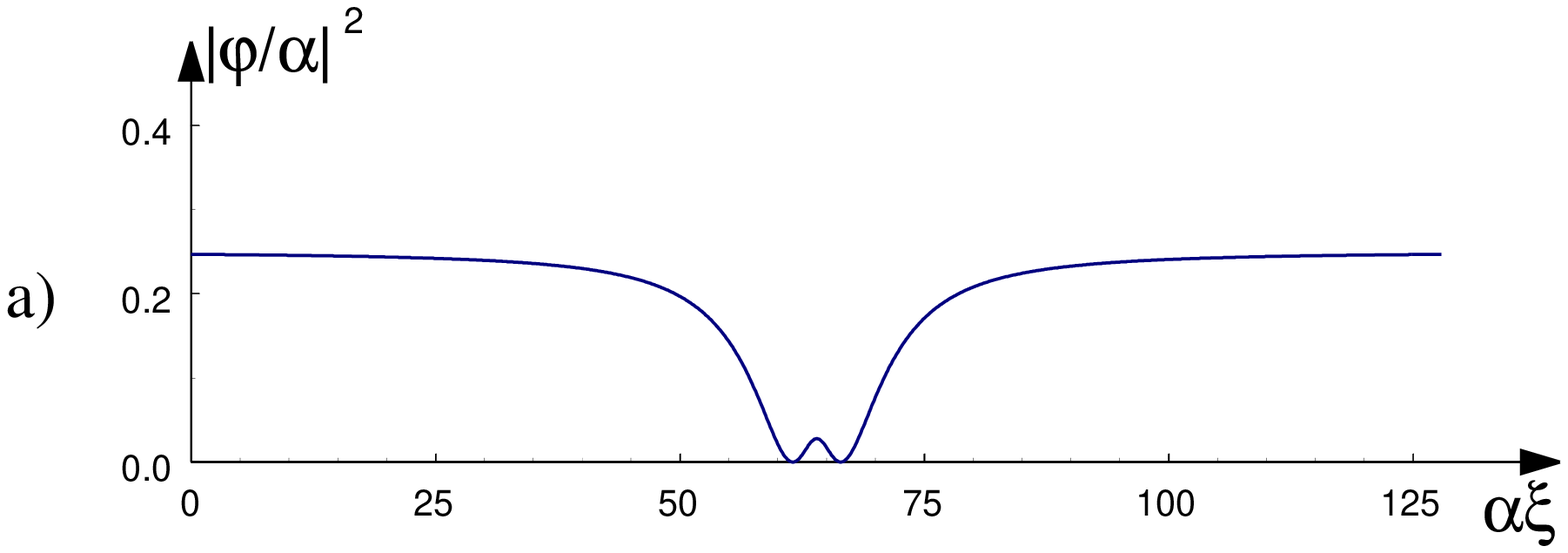,width=130mm}} %
\centerline{\psfig{figure=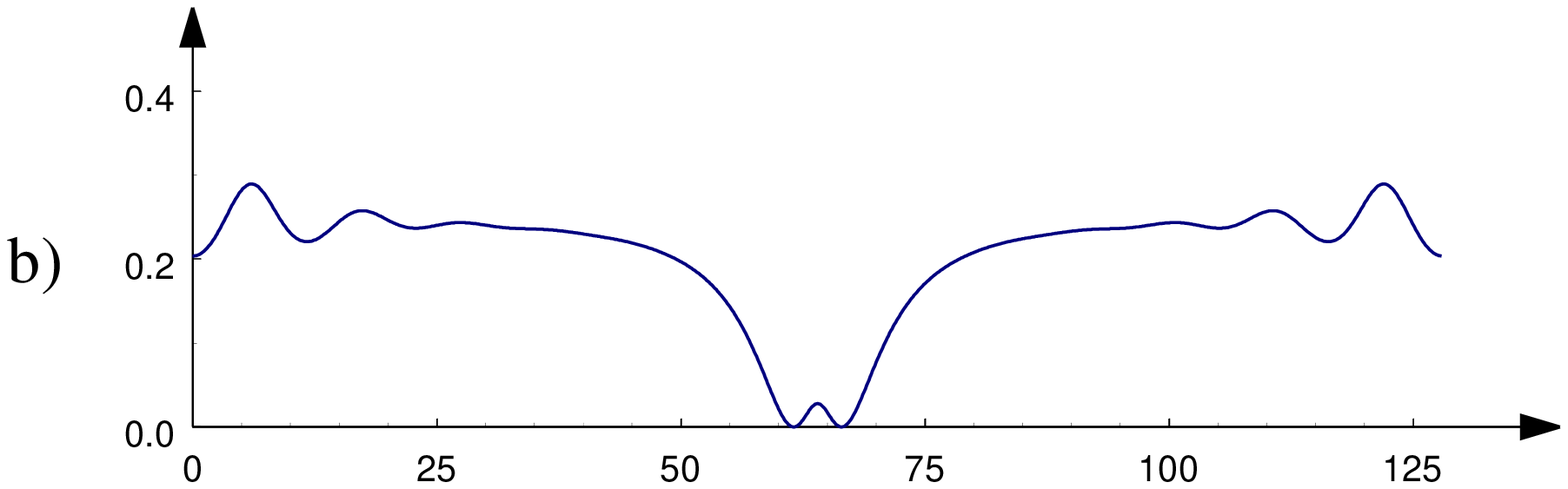,width=130mm}} %
\centerline{\psfig{figure=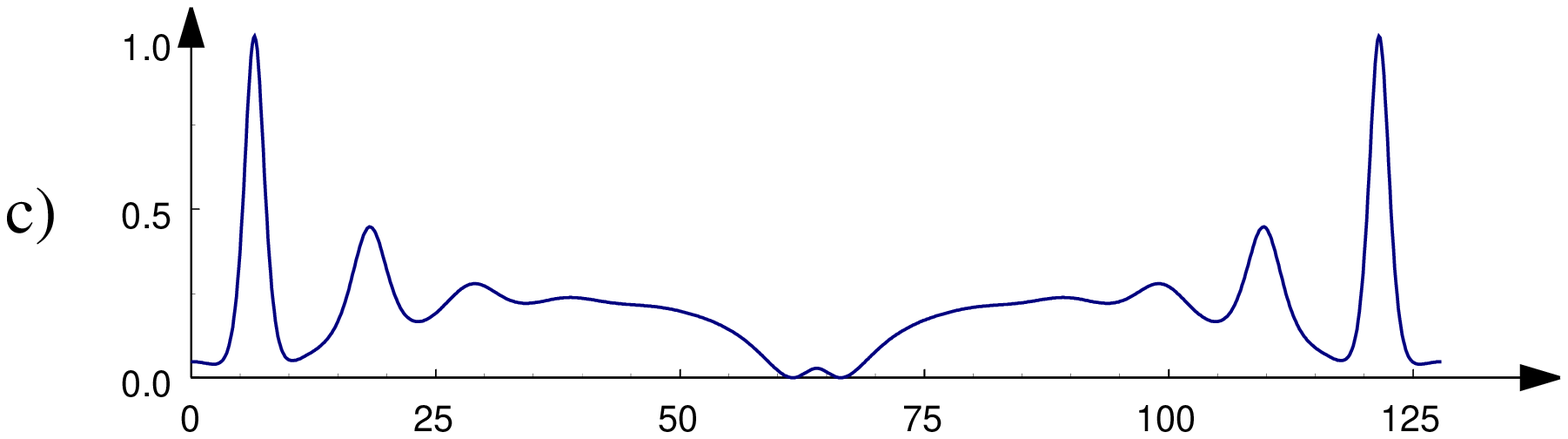,width=130mm}} %
\centerline{\psfig{figure=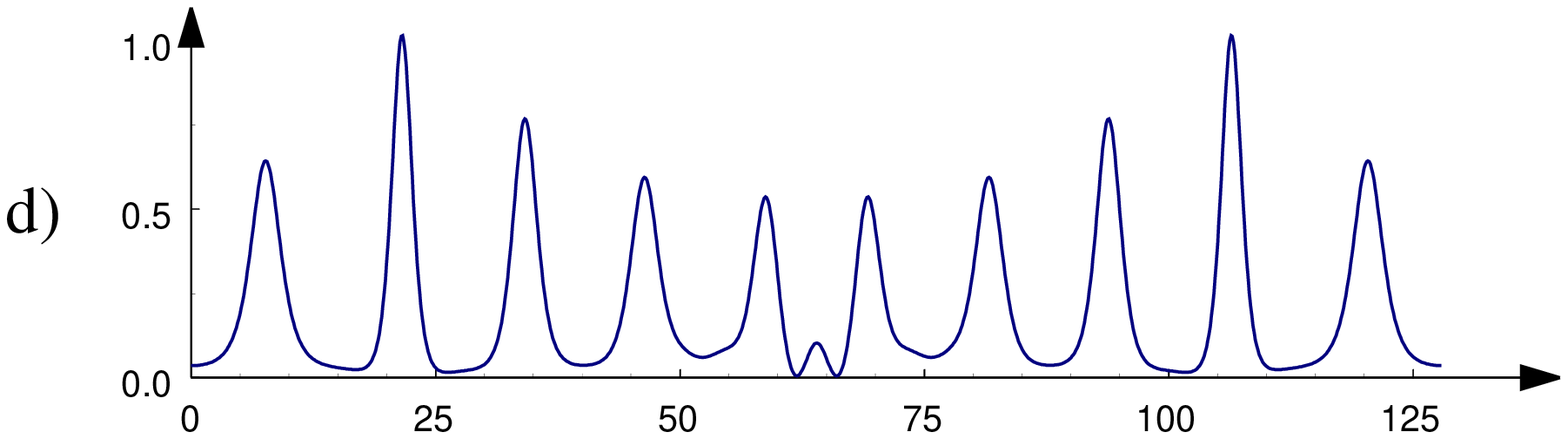,width=130mm}} %
\centerline{\psfig{figure=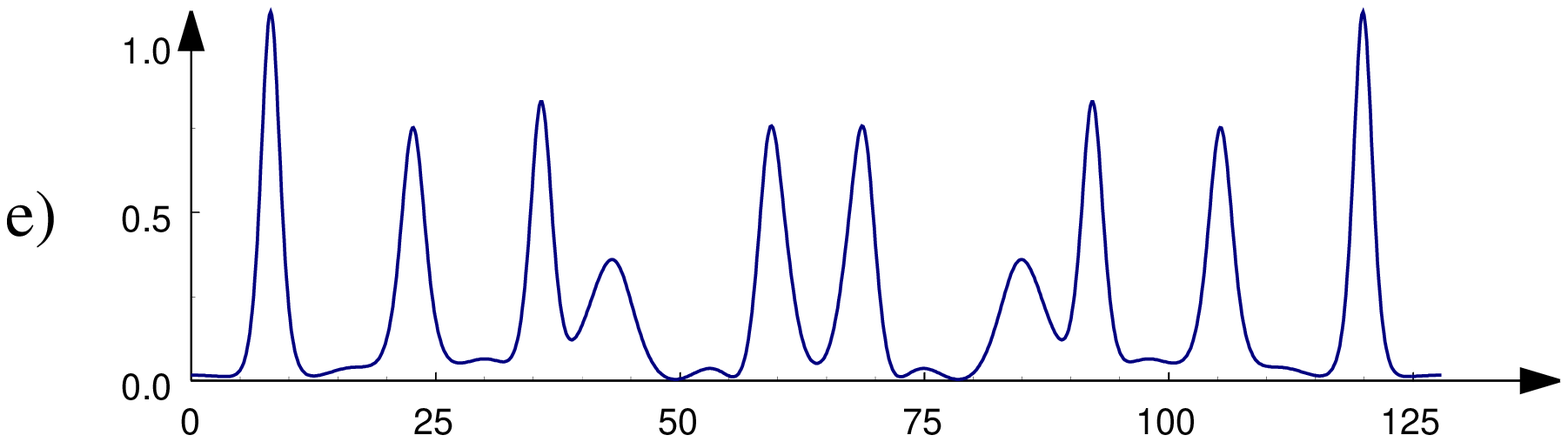,width=130mm}} %
\caption{An example of the instability development for the
algebraic soliton on a pedestal, given by Eq. (\protect\ref{AlgSonPed}) in
the model with the self-attraction: (a) -- $t=0$; (b) -- $t=30$; (c) -- $%
t=40 $; (d) -- $t=80$; (e) -- $t=100$ (the vertical scale in the three
latter cases is compressed by a factor of $2.5$).}
\label{f07old}
\end{figure}

\begin{figure}[h!]
\centerline{\psfig{figure=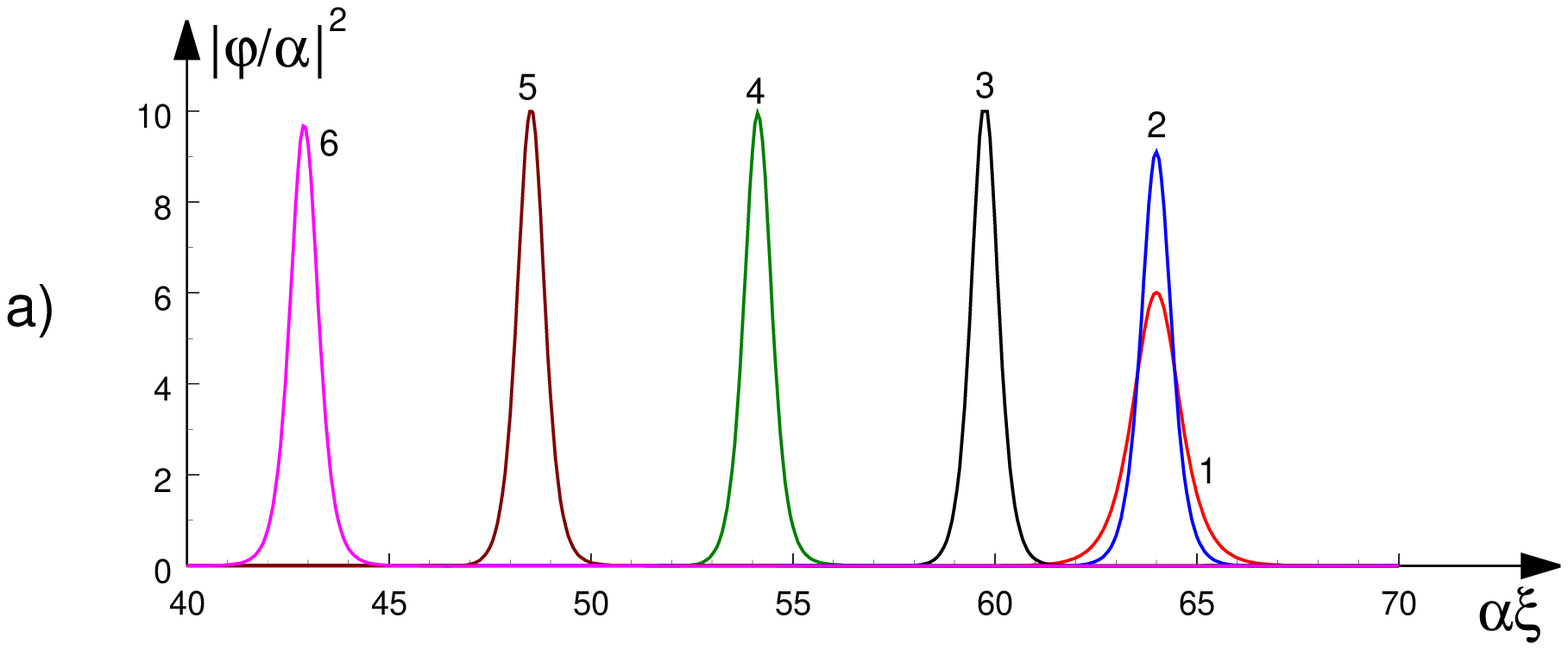,width=128mm}} \centerline{%
\psfig{figure=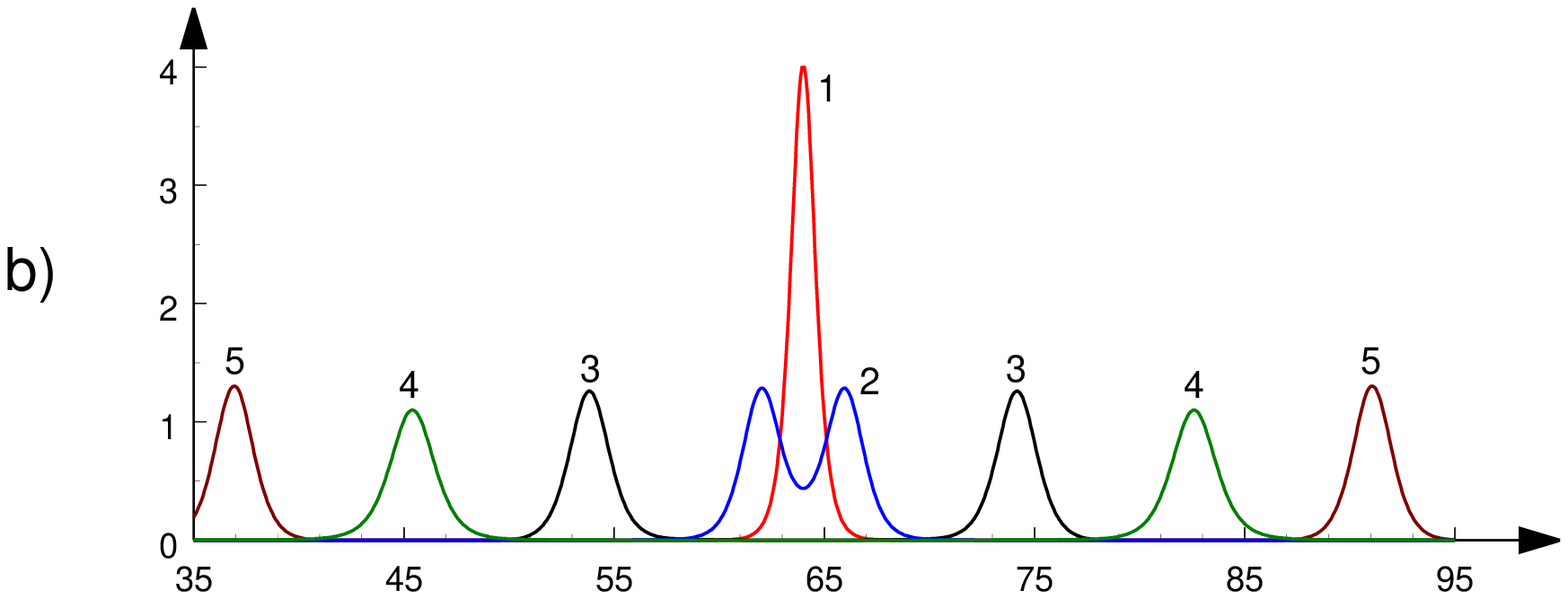,width=128mm}} \centerline{%
\psfig{figure=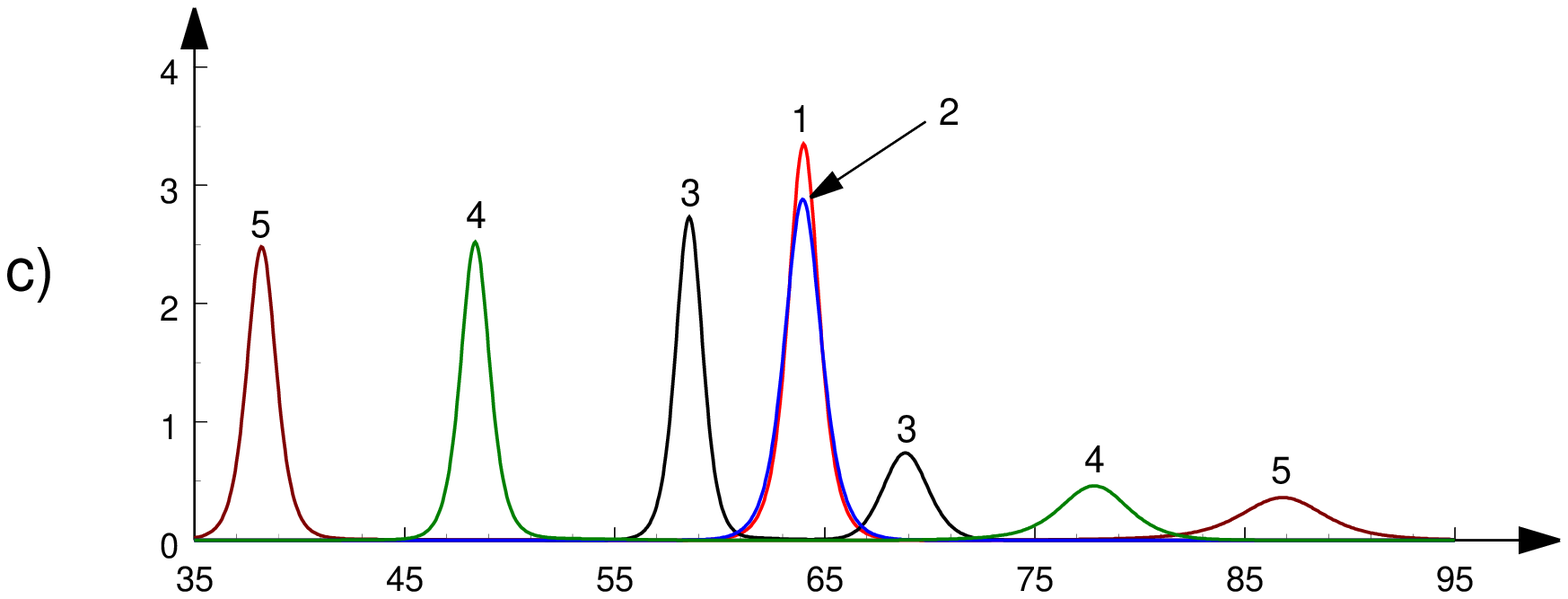,width=128mm}}
\caption{The evolution of a disturbed exponentially localized
soliton $\Phi _{-}$ from Eq. (\protect\ref{ExpSol}) (only a part of the
total spatial period of $L=128$ is shown). Panel (a) -- the initial
amplitude is $10\%$ greater than the stationary value corresponding to $%
\protect\alpha =\protect\nu =1$ (line 1 -- $t=0$, line 2 -- $t=12$, line 3
-- $t=16$, line 4 -- $t=18$, line 5 -- $t=20$, and line 6 -- $t=22$). Panel
(b) -- initial amplitude is $10\%$ smaller than the stationary value (line 1
-- $t=0$, line 2 -- $t=2$, line 3 -- $t=6$, line 4 -- $t=10$, and line 5 -- $%
t=14$). Panel (c) shows the evolution of initially undisturbed soliton (%
\protect\ref{ExpSol}), with $\protect\alpha =1$ and $\protect\nu =0.675$,
under the influence of small errors of the numerical truncation (line 1 -- $%
t=0$, line 2 -- $t=8$, line 3 -- $t=10$, line 4 -- $t=12$, and line 5 -- $%
t=14$).}
\label{f06}
\end{figure}

The stability of all the solutions found in the model with the
attractive nonlinearity was tested via direct simulations of Eq.
(\ref{dl1DGPE}). First, the expected instability of the solutions
on the pedestal and localized modes placed at the maximum of the
potential was corroborated. In the former case, the modulational
instability of the background leads to the formation of a chaotic
``gas" of interacting solitons. An examples of such instability onset is shown in Fig.~\ref{f07old}, which shows a spatial distribution of the density, $|\varphi (\xi )|^{2}$, at different moments of time when the spatial period is $L=128$. In the latter case, small random
perturbations may either cause the soliton to roll down from the
unstable position (see an example in Fig.~\ref{f06}a), or split --
symmetrically (see Fig.~\ref{f06}b), or sometimes asymmetrically
(see Fig.~\ref{f06}c) -- into two solitons moving in opposite directions. In particular,
the splitting was, naturally, observed under the action of an
initial perturbation which made the amplitude of the unstably
pinned soliton smaller, hence making it more similar to a
quasi-linear wave packet which is subject to the splitting by the
potential barrier. In fact, the strongly asymmetric splitting may
be realized as a result of strong emission of radiation from the
unstable soliton and self-retrapping of the emitted wave packet
into a small-amplitude soliton. Because the simulations were run
in the domain with periodic boundary conditions, we also observed
that, in the
case of the splitting of the unstable soliton into two, like in Figs.~\ref%
{f06}b and \ref{f06}c, the secondary solitons survived in the
course of numerous head-on collisions in the course of their
circular motion. If the instability gave rise to a moving soliton
and a radiation wave-train, the secondary interactions between
them would not destroy the soliton either.

\begin{figure}[h!]
\centerline{\psfig{figure=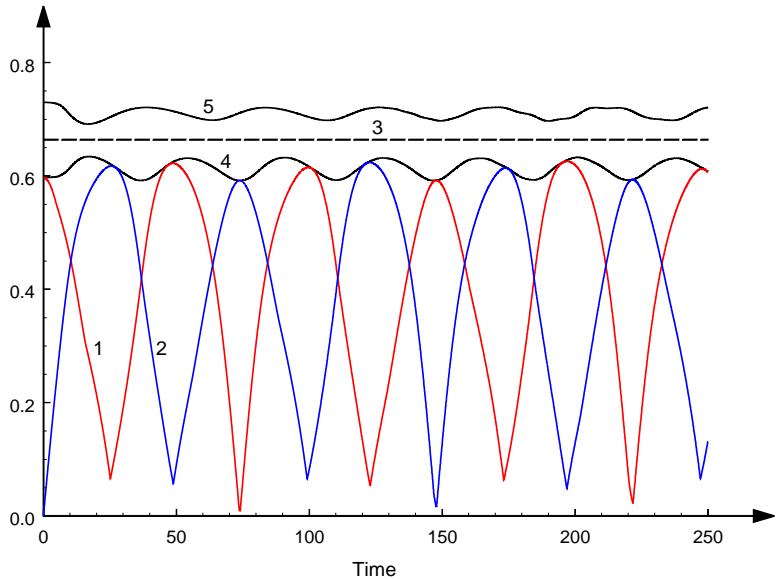,height=80mm}}
\caption{The evolution of perturbed solution $\Phi _{+}$ from
Eq. (\protect\ref{ExpSol}) in the model with the attractive nonlinearity,
when the amplitude of the initial perturbation was $10\%$ smaller than
needed for the stationary solution. Line 1 -- max$_{\protect\xi }|\mathrm{Re}%
\left\{ \protect\varphi (\protect\xi ,t)\right\} |$; line 2 -- max$_{\protect%
\xi }|\mathrm{Im}\left\{ \protect\varphi (\protect\xi ,t)\right\} |$.
Horizontal line 3 shows the constant amplitude of stationary soliton $\Phi
_{+}$ with $\protect\nu =1$; lines 4 and 5 show the time dependence of max$_{%
\protect\xi }|\protect\varphi (\protect\xi ,t)|$ in cases when the soliton's
amplitude was reduced (line 4) or increased (line 5) by $10\%$ against the
stationary value.}
\label{f07}
\end{figure}

Also in agreement with the expectation formulated above, the
numerical solutions demonstrate the \emph{stability} of solitons
$\Phi _{+}$ gives by
Eq.~(\ref{ExpSol}). In this case, the amplitude of the perturbed soliton, $%
\mathrm{\max }_{\xi }|\varphi (\xi )|$, varies in time around some
average value, as shown by lines 4 and 5 in Fig.~\ref{f07}, and
remains within the same range of the deviation from the \linebreak

\clearpage

\noindent%
stationary value as the initial perturbation. The stability of these solitons
is similar to that which was demonstrated above for the table-top
soliton in Fig.~\ref{f02} in the model with repulsive
nonlinearity.

\subsection{Other stationary solutions obtained from \textit{nonstationary}
solutions of the auxiliary Gardner equation}
\label{ss2.D}

The nonstationary version of GE (\ref{Gardner}),
\begin{equation}
\phi _{\tau }+\left( c\phi +\phi _{\xi \xi }-\phi ^{2}-\sigma \phi
^{3}\right) _{\xi }=0.  \label{NStGardner}
\end{equation}%
can also be used for the purpose of generating stable stationary solutions
to the GPE [note that here, in comparison with Eq. (\ref{Gardner}), we set $%
\alpha =-1$]. We stress that formal temporal variable $\tau $ in this
equation has nothing to do with physical time $t$ in Eq. (\ref{dl1DGPE}).

Equation (\ref{NStGardner}) is tantamount to the integrable modified KdV
equation \cite{AblSegur,Miles81,SlyunPel99,Pelinovsky-Grimshaw}. It may be
formally integrated once in $\xi $ and cast in the following form:
\begin{equation}
c\phi +\frac{\partial ^{2}\phi }{\partial \xi ^{2}}-\Phi (\xi ,\tau )\phi
-\sigma \phi ^{3}=-\left[ \frac{1}{\Phi (\xi ,t)}\frac{\partial }{\partial
\tau }\int_{0}^{\xi }\Phi (x^{\prime },\tau )\,dx^{\prime }\right] \phi ,
\label{NStGard}
\end{equation}%
where $\Phi (x,\tau )$ is one of the particular nonstationary solutions to
Eq. (\ref{NStGardner}). The term on the right-hand side of Eq.~(\ref{NStGard}%
) can be combined with term $\Phi (\xi ,\tau )\phi $ on the left-hand side,
to generate the \emph{stationary} GPE in the form of Eq. (\ref{St1DGPE})
with $c=\mu $ and potential $u(\xi ,\tau )$ which depends on \emph{free
parameter} $\tau $:
\begin{equation}
u(\xi ,\tau )=\Phi (\xi ,\tau )-\frac{1}{\Phi (\xi ,\tau )}\frac{\partial }{%
\partial \tau }\int_{0}^{\xi }\Phi (x^{\prime },\tau )\,dx^{\prime },
\label{EffPot}
\end{equation}%
cf. Eq. (\ref{u}) for the case when the stationary GE was used. It is worthy
to stress that, in the present case, the solution and the potential which
support it are no longer proportional to each other.

\begin{figure}[h]
\centerline{\psfig{figure=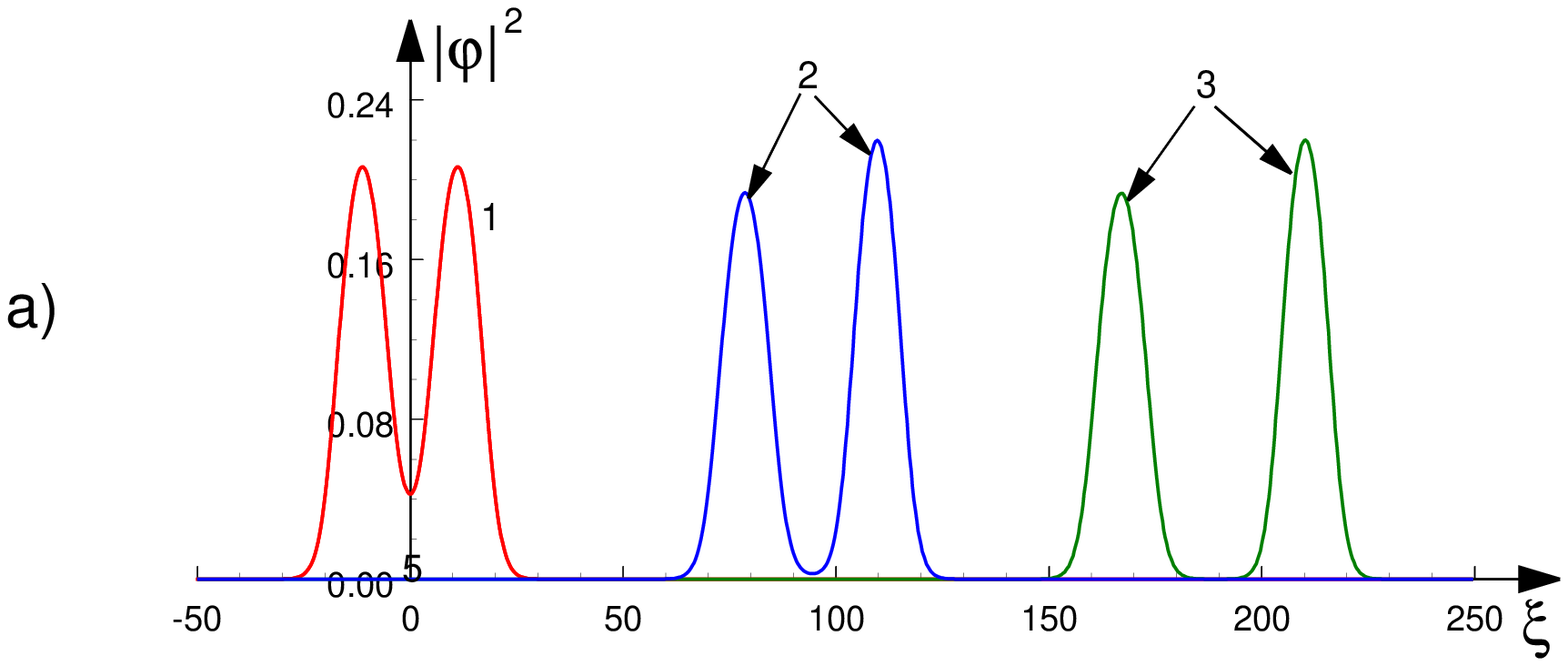,height=50mm}} \centerline{%
\psfig{figure=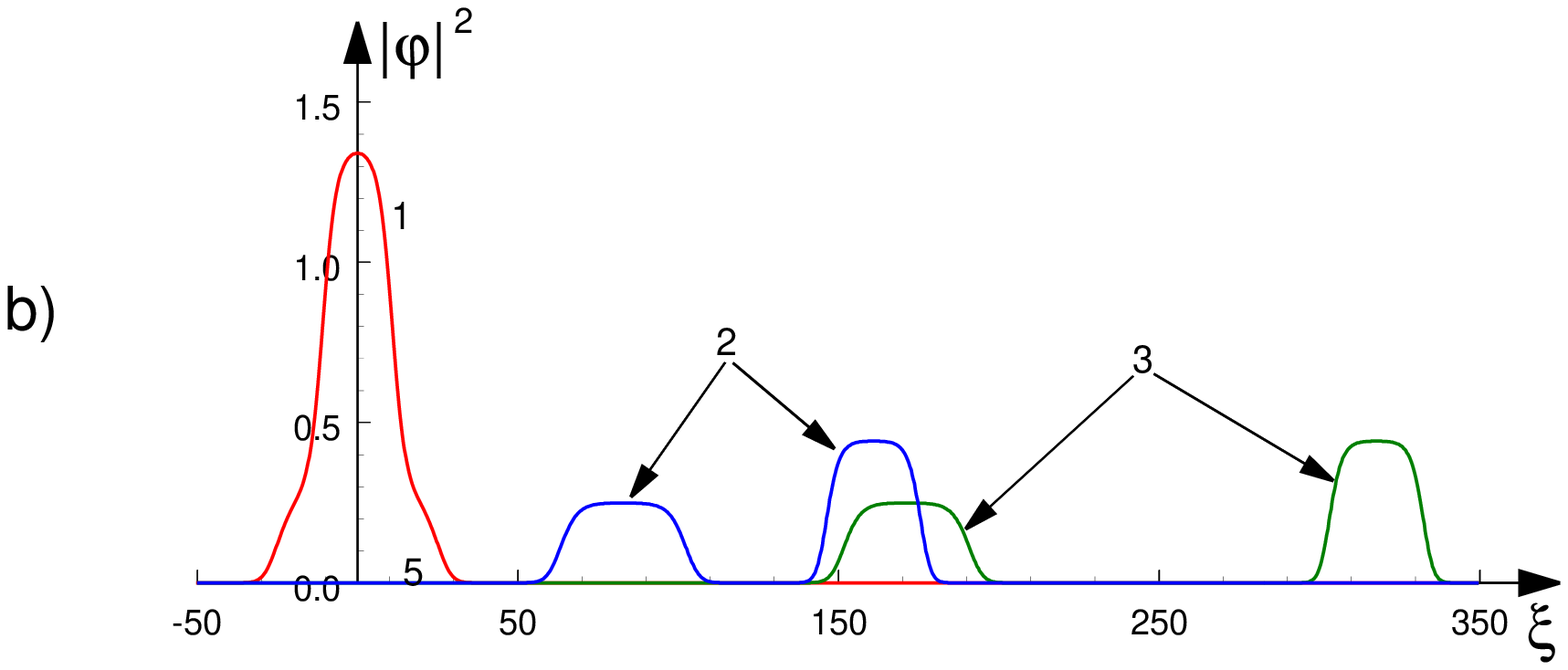,height=50mm}}
\caption{Solutions to the nonstationary Gardner equation (%
\protect\ref{NStGardner}) with $c=0$ and $\protect\sigma =1$, which describe
disintegration of the initial configuration into a pair of ``fat solitons".
They emerge with parameters $\protect\nu _{1}=0.75$, $\protect\nu _{2}=0.8$
in (a), and $\protect\nu _{1}=0.75$, $\protect\nu _{2}=0.999$ in (b). Lines
1, 2, and 3 correspond, respectively, to $\protect\tau =0$, $\protect\tau %
=1500,$ and $\protect\tau =3000$.}
\label{f08}
\end{figure}

As an example, we take, following Ref. \cite{SlyunPel99}, a solution to
nonstationary GE (\ref{NStGardner}) with $\sigma =1$, which describes
disintegration of the initial configuration into two fat solitons:
\begin{equation}
\Phi (\xi ,\tau )=\frac{\nu _{1}^{2}-\nu _{2}^{2}}{3}\left( \frac{1}{%
Z_{2+}-Z_{1+}}-\frac{1}{Z_{2-}-Z_{1-}}\right) ,  \label{TwoSolGardner}
\end{equation}%
where
\begin{equation}
Z_{1\pm }\equiv \nu _{1}\tanh {\left\{ \frac{\nu _{1}\sqrt{2}}{6}\left[ \xi
-\left(-\mu +\frac{2\sqrt{2}}{27}\nu _{1}^{2}\right) \tau \pm \delta _{1}%
\right] \right\} },  \label{Z1}
\end{equation}%
\begin{equation}
Z_{2\pm }\equiv \nu _{2}\coth {\left\{ \frac{\nu _{2}\sqrt{2}}{6}\left[ \xi
-\left(-\mu +\frac{2\sqrt{2}}{27}\nu _{2}^{2}\right) \tau \pm \delta _{2}%
\right] \right\} },  \label{Z2}
\end{equation}%
\begin{equation}
\delta _{1,2}=\left( 2\nu _{1,2}\right) ^{-1}\ln \left[ \left( 1+\nu
_{1,2}\right) /\left( 1-\nu _{1,2}\right) \right] .  \label{delt}
\end{equation}%
Two examples of these solutions are shown in Fig.~\ref{f08} for a fixed
value of $\nu _{1}=0.75$ and two different values of the other parameter, $%
\nu _{2}=0.8$ and $0.999$. The configuration shown in Fig.~\ref{f08}(b)
actually represents a pair of table-top Gardner solitons (\ref{FatSol}) with
different parameters. Solution (\ref{TwoSolGardner})--(\ref{delt}) may be
treated as a continuous family (parameterized by $\tau $) of stationary
solutions to Eq.~(\ref{St1DGPE}), with the corresponding potential produced
by Eqs.~(\ref{TwoSolGardner})--(\ref{delt}).

To conclude this subsection, it is relevant to mention exact nonstationary
solutions in the form of breathers, for Gardner equation (\ref{NStGardner})
corresponding to the GPE with the attractive nonlinearity ($\sigma =-1$)
\cite{Pelinovsky-Grimshaw}. Such breathers look as two periodically
interacting exponentially localized solitons, or as envelope solitons of the
NLSE. At any fixed value of $\tau $ in Eq.~(\ref{NStGardner}), the breathers
generate, as outlined above, exact stationary solutions of the GPE with the
corresponding potentials given by Eq.~(\ref{EffPot}).

\subsection{Reconstruction of the supporting potential in the GPE for an
arbitrary matter-wave distribution}

\label{ss2.E}

An arbitrary distribution of the stationary matter-wave field, $\varphi(\xi)$%
, can be made an exact solution to stationary GPE (\ref{St1DGPE}), if the
potential in the equation is chosen as

\begin{equation}
u(\xi ) = \frac{\varphi ^{\prime \prime }}{\varphi (\xi )} - \sigma
\varphi ^{2}(\xi )+\mu .  \label{InversePot}
\end{equation}%
Below, we demonstrate that this, seemingly ``trivial", approach may also
produce essential results.

\begin{figure}[t]
\vspace*{-20mm}
\begin{tabular}{cc}
\hspace*{-10mm}\psfig{figure=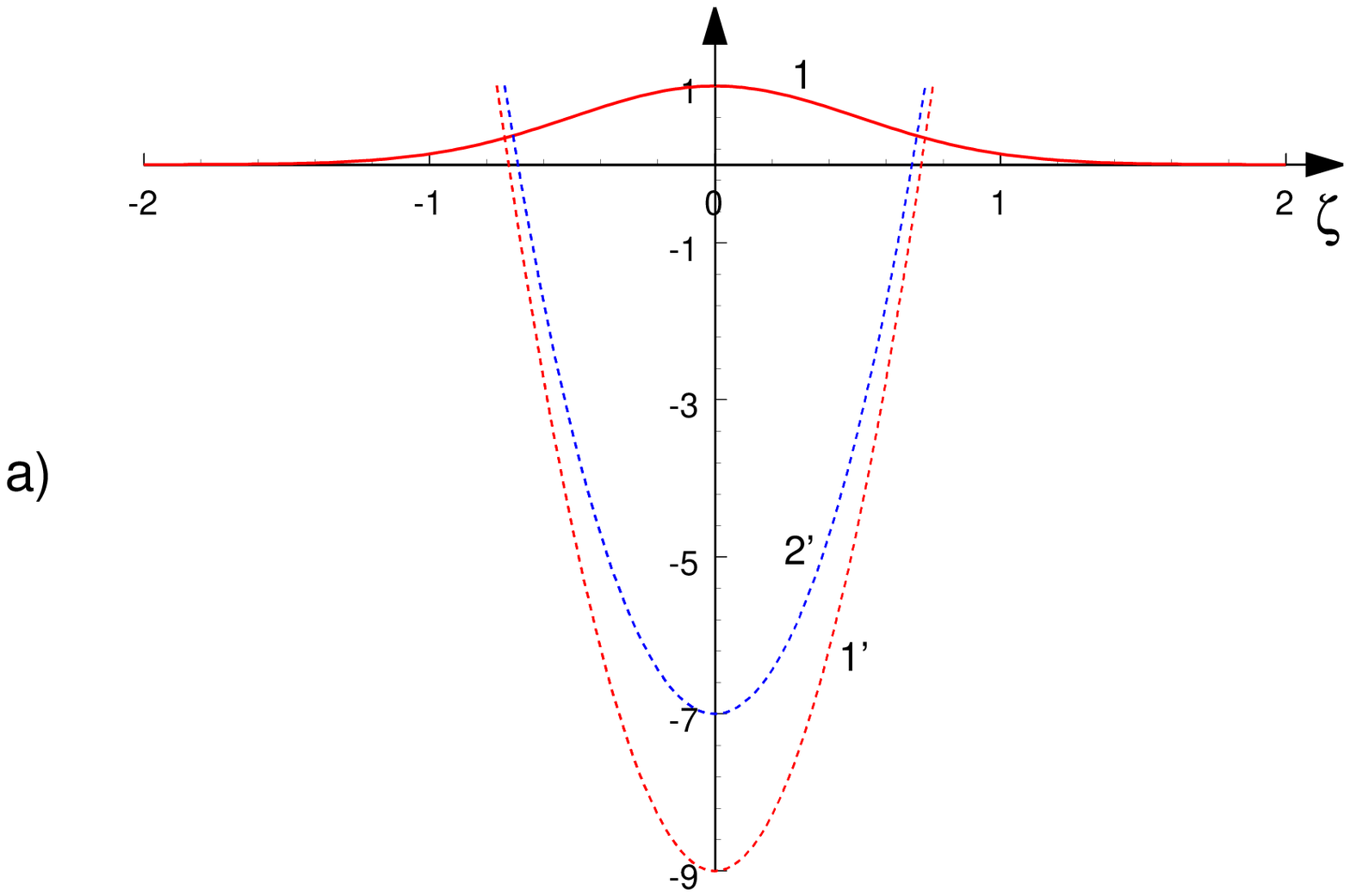,height=55mm} & \psfig{figure=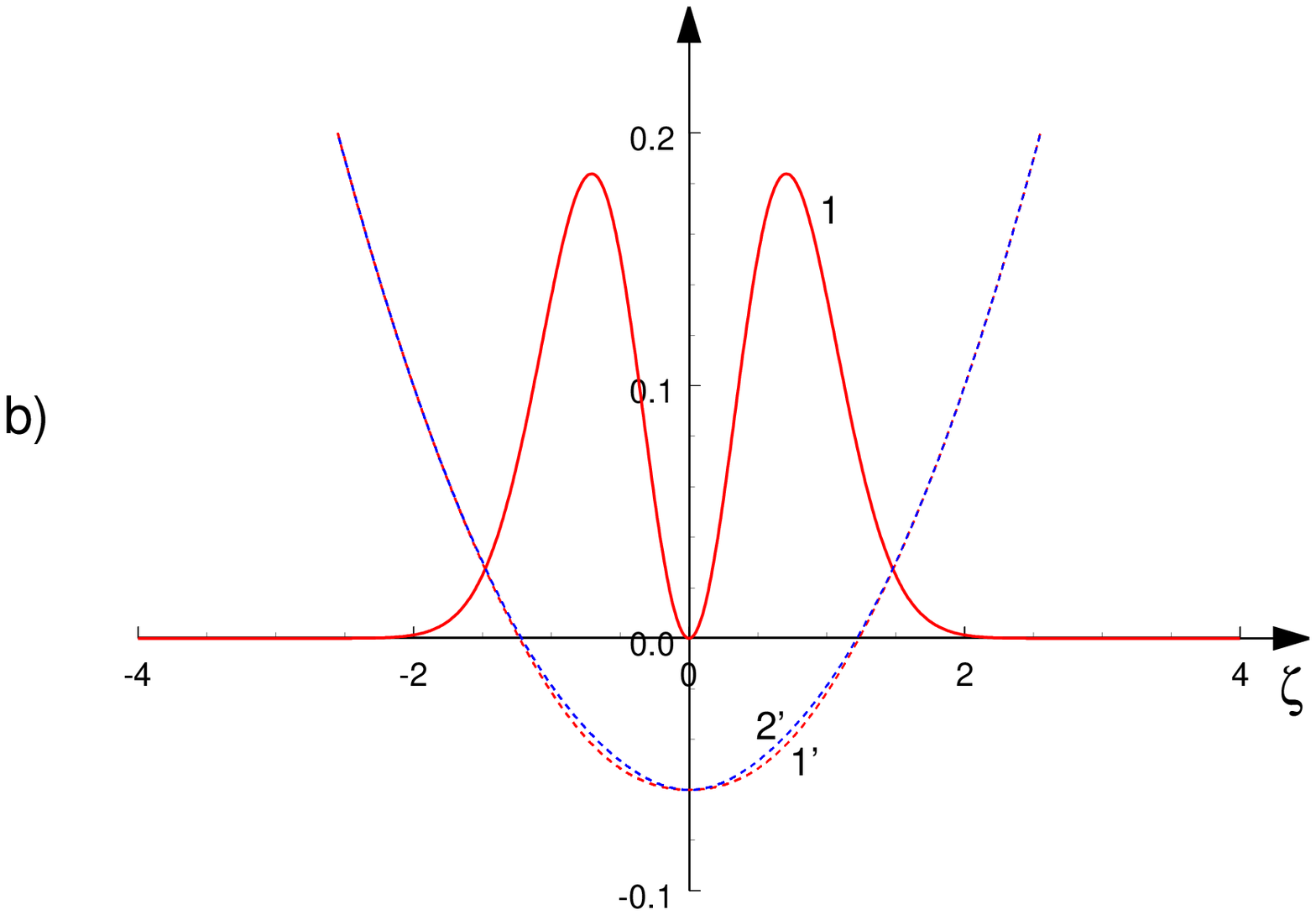,height=55mm}
\\
\hspace*{-10mm}\psfig{figure=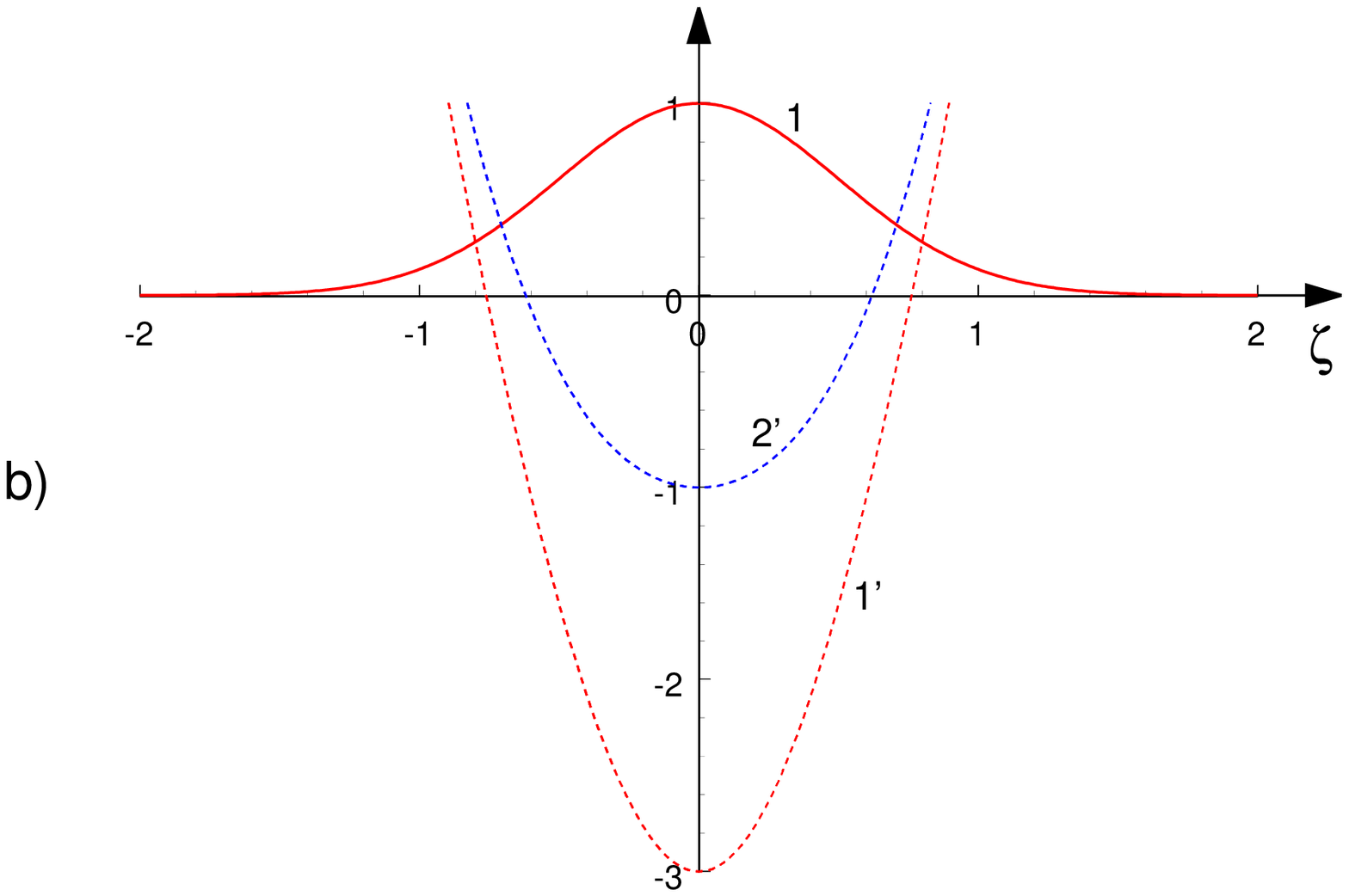,height=55mm} & \psfig{figure=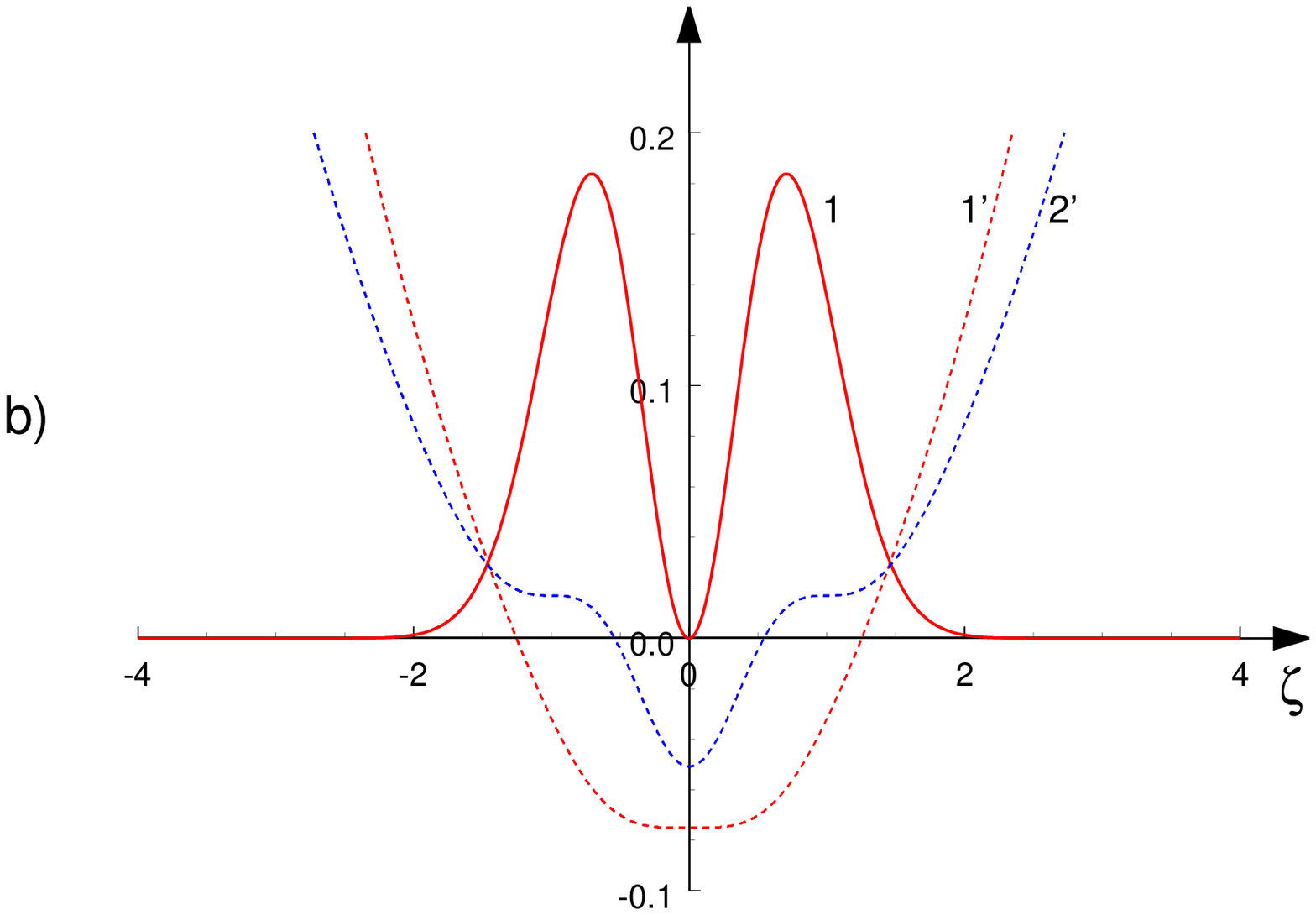,height=55mm}
\\
\hspace*{-10mm}\psfig{figure=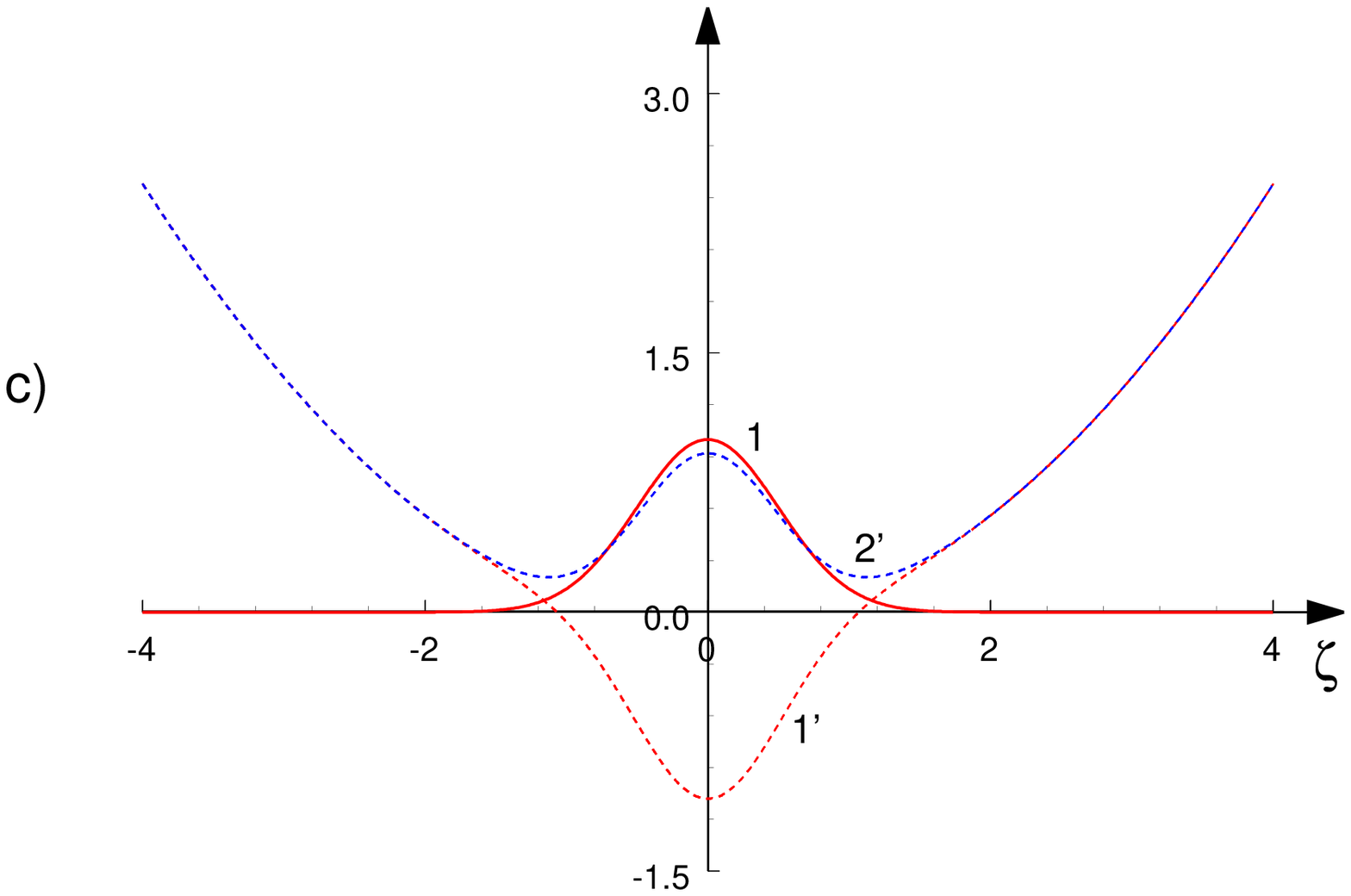,height=55mm} & \psfig{figure=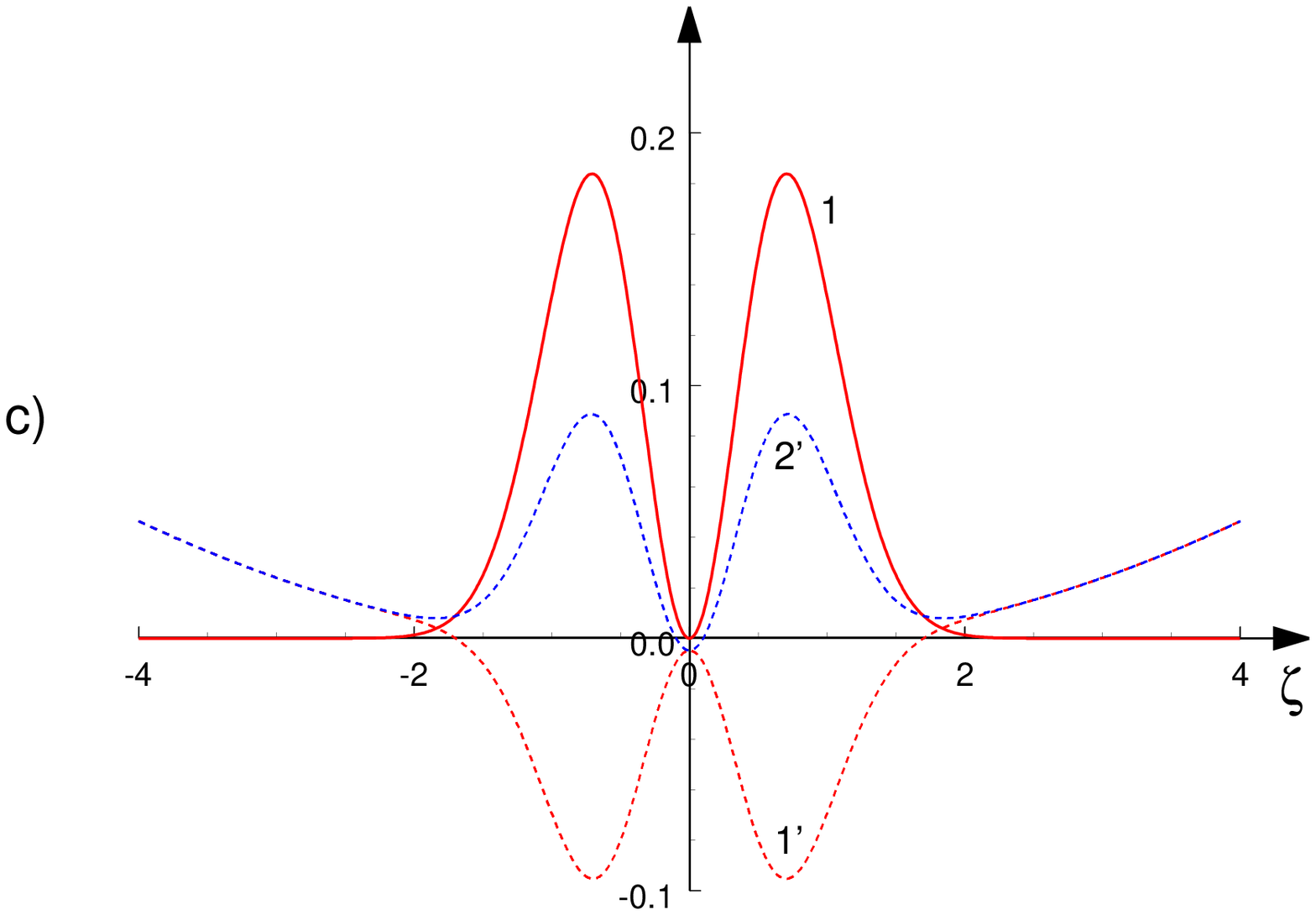,height=55mm}
\\
FIG.~11 & FIG.~12%
\end{tabular}%
\caption{Normalized Gaussian solution (\protect\ref{InvPotS1}%
) in terms of $F^{2}(\protect\zeta )$ (solid curve $1$) and
corresponding normalized potentials $v_{e}(\protect\zeta )$
(broken curves $1^{\prime }$
and $2^{\prime }$, pertaining to $\protect\sigma =1$ and $\protect\sigma =-1$%
, respectively), as functions of normalized coordinate
$\protect\zeta $. Panels a), b), and c) were generated for
$S=0.5$, $S=1$, and $S=5$, respectively. Note that the plots are
shown on different scales.}
\label{f09}%
\caption{The same as in Fig.~\protect\ref{f09} but
for the derivative-Gaussian solution (\protect\ref{InvPotS2}). The
corresponding
potentials are additionally reduced by the factor $K$: $v_{e}(\protect\zeta %
)/K$, where $K$ is different in each panel. Panel a): $S=1$, $K = 100$ for $%
\protect\sigma = \pm 1$; panel b): $S=2$, $K = 20$ for $\protect\sigma = 1$
and $S=2e$, $K = 4$ for $\protect\sigma = -1$; and panel c): $S=25$, $K = 2$
for $\protect\sigma = \pm 1$.}
\label{f10}
\end{figure}

\label{sss2.E.1}\textit{An example: the Gaussian profile of the matter wave}%
. First, we take a Gaussian matter-wave pulse, which is the case of obvious
interest to applications, $\varphi (\xi )=A\exp {[-(\xi /l)^{2}]}$.
Substituting this into Eq.~(\ref{InversePot}), one finds:
$$
u(\xi ) = \frac{2}{l^{2}}\left( 2\frac{\xi ^{2}}{l^{2}} - 1\right) -\sigma A^{2}\exp {[-2(\xi
/l)^{2}]}+\mu .
$$
The trial solution, $\varphi (\xi )$, and the corresponding
potential can be presented in the dimensionless form:
\begin{equation}
F(\zeta )=\exp {(-\zeta ^{2})}, \quad v_{e}(\zeta ) = \frac{2}{S^{2}}\left( 2\zeta ^{2}-1\right) -\sigma \exp {(-2\zeta ^{2})} + M,
\label{InvPotS1}
\end{equation}%
where $\zeta =\xi /l$, $F(\zeta )=\varphi (\zeta )/A$, $v_{e}(\zeta
)=u(\zeta )/A^{2}$, $S=Al$, $M = \mu /A^{2}$. While $F(\zeta )$ does not
contain any parameter, the dimensionless potential depends on two
independent constants, $S$ and $M$, for each sign of $\sigma =\pm 1$. Figure %
\ref{f09} shows the squared normalized solution, $F^{2}(\zeta )$, and the
corresponding potentials for both signs of $\sigma $, as given by Eq.~(\ref%
{InvPotS1}) for several values of $S$ and $M=0$. Note that, in the case
displayed in Fig. \ref{f09}c, the potential corresponding to the GPE with
the self-attraction nonlinearity ($\sigma = -1$) features a \emph{double-well%
} shape. As follows from Eq. (\ref{InvPotS1}), such a shape may occurs only
in the case of $\sigma = -1$, provided that $S$ exceeds a threshold value, $%
S_{\mathrm{thr}} = \sqrt{2}$.

\label{sss2.E.2}\textit{A derivative-Gaussian profile of the
matter wave}. Trial function $\varphi (\xi )$ used above is an
even function of $\xi $ without nodes, which, apparently,
represents the ground state for the nonlinear GPE with the given
potential. Here we aim to consider another example, when the trial
function is chosen as an odd one, with a single
node, thus representing the first excited state. To this end, we take $%
\varphi (\xi )=A\xi \exp \left[ {-}\left( {\xi /l}\right)
{^{2}}\right] $ and derive the corresponding potential from
Eq.~(\ref{InversePot}):
$$
u(\xi ) = \frac{ 2}{l^{2}}\left( 2\frac{\xi^{2}}{l^{2}} - 3\right) -\sigma A^{2}\xi ^{2}\exp {[-2(\xi
/l)^{2}]}+\mu.
$$
It is again convenient to present the trial
solution, $\varphi (\xi )$, and the corresponding potential in the
dimensionless form:
\begin{equation}
F(\zeta )=-\zeta \exp {(-\zeta ^{2})}, \quad v_{e}(\zeta ) = \frac{2}{S^{2}} \left( 2\zeta ^{2}-3\right) -\sigma \zeta ^{2}\exp
{(-2\zeta ^{2})}+M,  \label{InvPotS2}
\end{equation}%
where, this time, $\zeta =\xi /l$, $F(\zeta )=\varphi (\zeta )/A$, $%
v_{e}(\zeta )=u(\zeta )/A^{2}l^{4}$, $S=Al^{2}$, and $M=\mu
/(Al)^{2}$. Plots corresponding to this trial function and the
supporting potentials are
displayed in Fig.~\ref{f10} for $M=0$ and both signs of the parameter $%
\sigma $. Apparently, function $F(\zeta )$ represents the first
excited
eigenmode in the corresponding potential well $%
v_{e}(\zeta )$. Note that, as it follows from
Eq.~(\ref{InvPotS2}), the
trapping potential has a single-well structure at $S \le S_{\mathrm{thr}%
}^{+} \equiv 2$ for $\sigma = +1$, and at $S \le
S_{\mathrm{thr}}^{-}\equiv 2e$ for $\sigma = -1$. In the opposite
case, the potential acquires the double-well structure when $S >
S_{\mathrm{thr}}^{+}$ for $\sigma = +1$ (see line $1^{\prime}$ in
Fig.~\ref{f10}c), and the triple-well structure when $S
> S_{\mathrm{thr}}^{-}$ for $\sigma = -1$ (see line $2^{\prime}$ in Fig.~\ref%
{f10}c). The shapes of the potential at the critical values of $S=S_{\mathrm{%
thr}}^{\pm}$ are shown in Fig.~\ref{f10}b.

\textit{A comb-top-Gaussian profile of the mater wave}. Here we consider the
trial function in the form of the Gaussian with a superimposed ``comb'',
which corresponds to the physically relevant combination of an OL and
external parabolic trap:
\begin{equation}
\varphi (\xi )=A\exp {(-\xi ^{2}/l_{1}^{2})}+B\cos \left( {kx}\right) \exp {%
(-\xi ^{2}/l_{2}^{2})},  \label{CombGauss}
\end{equation}%

\begin{figure}[t]
\vspace*{-20mm}
\centerline{\psfig{figure=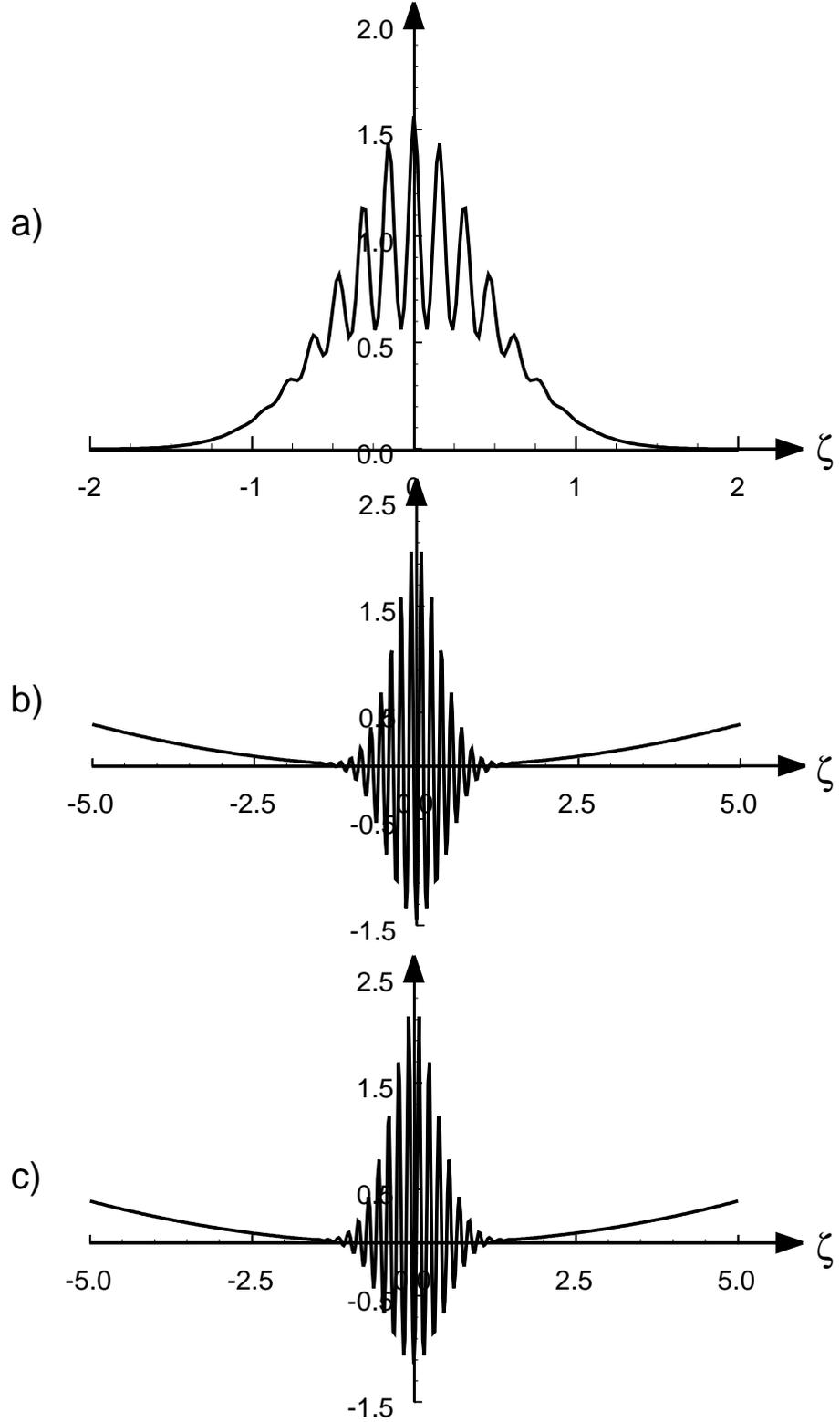,height=210mm}}
\caption{Squared solution (\protect\ref{NormCombGauss}), $F^{2}(\protect%
\zeta )$, (panel a) and the corresponding potential (\protect\ref%
{CombGaussPot}) (reduced by the factor of $10$) for $\protect\sigma =1$
(panel b) and $\protect\sigma =-1$ (panel c) as functions of $\protect\zeta $. Parameters are $S=5$, $M=0$, $%
\protect\kappa =40$, $b=0.25$, $\protect\varepsilon =2$.}
\label{f11}
\end{figure}

\clearpage

\noindent%
where $A$, $B$ and $l_{1,2}$ are arbitrary constants. This function
resembles, in particular, a nu\-me\-rical solution which was found in Ref.
\cite{Chiofalo00} (see also review \cite{Minguzzi04}) for the OL potential.
Function (\ref{CombGauss}) and the corresponding potential, as given by Eq.~(%
\ref{InversePot}), can be presented in the following dimensionless form:
\begin{equation}
F(\zeta )=\exp {(-\zeta ^{2})}+b\cos \left( {\kappa \zeta }\right) \exp {%
[-(\varepsilon \zeta )^{2}]},  \label{NormCombGauss}
\end{equation}%
\begin{eqnarray}
v_{e}(\zeta ) &=&\frac{4}{S^{2}}\frac{\zeta ^{2}-\frac 12+b\left[ \left(
\varepsilon ^{4}\zeta ^{2}- \frac{\varepsilon ^{2}}{2} - \frac{\kappa ^{2}}{4}\right) \cos
\left( {\kappa \zeta }\right) +\varepsilon ^{2}\kappa \zeta \sin \left( {%
\kappa \zeta }\right) \right] \exp {[-(\varepsilon ^{2}-1)\zeta ^{2}]}}{%
1+b\cos \left( {\kappa \zeta }\right) \exp {[-(\varepsilon ^{2}-1)\zeta ^{2}]%
}}{}  \notag \\
{} && \nonumber \\
{} &&{}{}-\sigma \exp {(-2\zeta ^{2})}\left\{ 1+b\cos \left( {\kappa \zeta }%
\right) \exp {\left[ -(\varepsilon ^{2}-1)\zeta ^{2}\right] }\right\} ^{2}-M,
\label{CombGaussPot}
\end{eqnarray}%
where $F(\zeta )=\varphi (\zeta )/A$, $v_{e}(\xi )=u(\xi )/A^{2}$, and $%
\zeta =\xi /l_{1}$, $b=B/A$, $\kappa =kl_{1}$, $\varepsilon =l_{1}/l_{2}$, $%
S=Al_{1}$, $M=-\mu /A^{2}$. Varying parameters $S$, $M$, $\kappa $, $b$ and $%
\varepsilon $, one can obtain a wide class of solutions. The corresponding
potentials asymptotically approach the parabolic shape at large $|\xi |$,
featuring a complex oscillatory shape at the center. Solution (\ref%
{NormCombGauss}), $F^{2}(\zeta )$ and the corresponding potential (\ref%
{CombGaussPot}) are shown in Fig.~\ref{f11} for both signs of $\sigma$. \\

\textit{The stability of the Gaussian-type solutions}. Stability
of all
solutions presented in this section was tested via simulations of Eq. (\ref%
{dl1DGPE}). The results are summarized as follows.

(1) In the case of the repulsive nonlinearity, $\sigma =1$, the
Gaussian
solution with potential (\ref{InvPotS1}) is \emph{stable} for all values of $%
S$. It is stable too in the case of $\sigma =-1$ (the attractive
nonlinearity) if the corresponding potential features the
single-well shape
(such as shown in Figs.~\ref{f09}a and 9b), i.e. $S\leq S_{\mathrm{thr}%
}\equiv \sqrt{2}$, see above. However, in the model with the
attractive nonlinearity, the solution naturally becomes unstable
when the single-well potential transforms into the double-well
potential, i.e. when $S > \sqrt{2}$ (see line $2^{\prime}$ in
Fig.~\ref{f09}c). In the latter case, the solution preserves its
shape until $t\lesssim 20$, and then spontaneously splits into two
pulses which quasi-regularly oscillate relative to each other.

\begin{figure}[t!]
\centerline{\psfig{figure=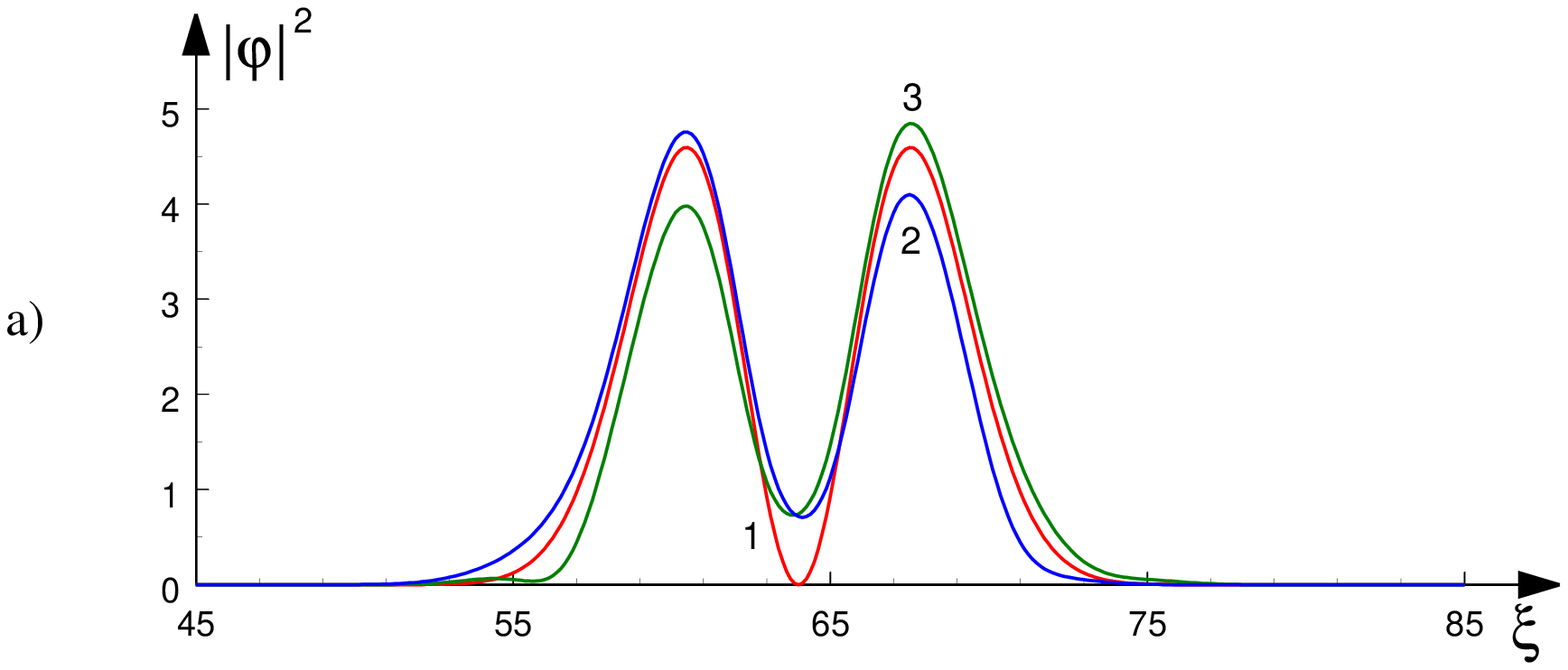,height=70mm}}
\centerline{\psfig{figure=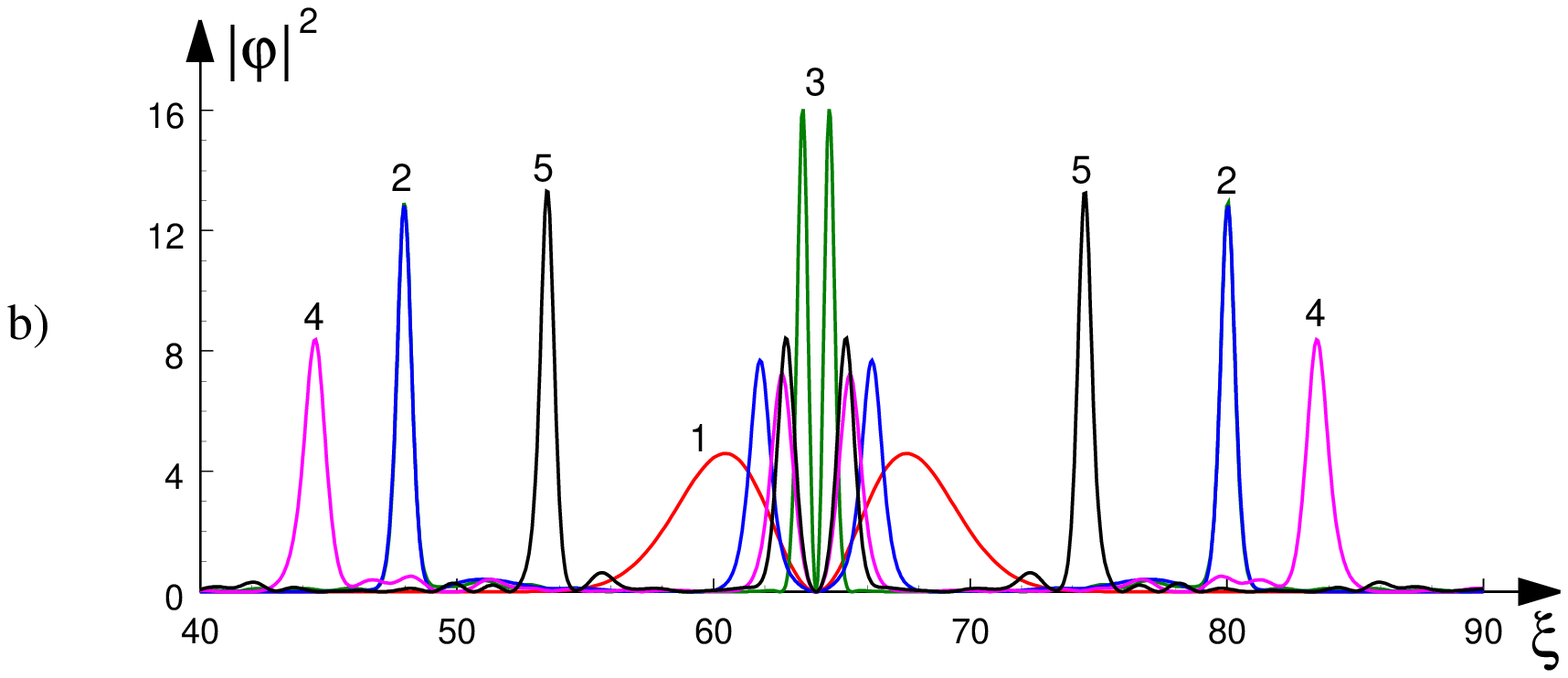,height=70mm}}
%\centerline{\psfig{figure=fig12-New.eps,height=70mm}}
\caption{The spatial profile of $|\protect\varphi
|^{2}$ at different times, for the derivative-Gaussian initial
configuration with $S=25$ (the size of the computation domain is $L=128$). In panel (a) pertaining to the case of $\protect\sigma =1$ lines 1, 2 and 3 correspond, respectively, to $t=0,$ 50 and 60. In panel (b) pertaining to the case of $\protect\sigma =-1$ lines 1, 2, 3, 4 and 5 correspond, respectively, to $t=0,$ 1, 2, 3, and 4 (note that
there are two pulses of smaller amplitudes near the center at
$t=1,$ 3 and 4).} \label{f13}
\end{figure}

(2) The derivative-Gaussian solution with potential (\ref{InvPotS2}) is
\emph{stable} in \emph{both cases} of $\sigma =\pm 1$, provided that the
underlying potential keeps the single-well shape, i.e. until $S \leq S^{+}_{%
\mathrm{thr}} \equiv 2$ for $\sigma = +1$ and $S \leq S^{-}_{\mathrm{thr}}
\equiv 2e$ for $\sigma = -1$ (see Figs.~\ref{f10}a and \ref{f10}b). At
greater values of $S$, the solutions in the double-well potential (see Fig.~%
\ref{f10}c) are unstable, for either sign of $\sigma$. However,
manifestations of the instability are different for $\sigma = +1$ and $%
\sigma = -1$. In the former case, the initial distribution was preserved in
a quasi-stable state until $t\approx 40 $. Then, the profile of $|\varphi
(\xi )|$ became asymmetric, with one maximum
(for instance, the right one, as in Fig.~\ref{f13}a)
being greater than the other. After reaching a well-pronounced asymmetric
shape, the process reverted, making the left maximum greater than the right
one. This process repeated persistently, so that the initial soliton \linebreak

\clearpage

%\clearpage

\noindent%
was eventually transformed into an immobile breather consisting of two
non-stationary pulses which oscillate in time quasi-randomly due to the
energy exchange between them and, apparently, due to their interaction with
a linear wave-train shed off by the pulses. It is worthy to note that the
instability of the odd mode trapped in the double-well potential shown, for
instance, in Fig.~\ref{f10}c, sets in via the breaking of the skew symmetry
of this mode, in agreement with the general principle that the repulsive
nonlinearity gives rise to the symmetry breaking of odd modes trapped in
double-well potentials \cite{DWP}.

In the case of $\sigma =-1$, the character of instability observed at $S >
S^{-}_{\mathrm{thr}}$ is different. For instance, at $S=25$ the solution
became unstable at $t\approx 5$, splitting into four pulses and a
small-amplitude wave train (see line 2 in Fig.~\ref{f13}). Two of those four
pulses moved to the left and two others -- to the right. After passing some
distance, these pulses bounced back from the parabolic potential, moved back
towards the center, and formed two very narrow and closely located spikes
(see line 3 in Fig.~\ref{f13}). Such cycles of the splitting and partial
recombination with the pulse compression in the vicinity of the center
repeated indefinitely long. Thus, some type of a double breather is formed
in this case too.

(3) The comb-top Gaussian solution with potential (\ref{CombGaussPot}) is
stable in the model with the repulsive nonlinearity, $\sigma =1$ (we did not
explore the stability of this solution in a wide range of parameters; here
we report an example for $b=1/4$, $\varepsilon =2$, $\kappa =40$, $M=0$, and
$S=5$). Namely, if external perturbations were added to the solution, e.g., $%
A$ in Eq.~(\ref{CombGauss}) was taken $10\%$ smaller or grater against its
stationary value, the evolution lead to time variations of $|\varphi |$
within the same $10\%$ range, similar to what was reported in item (1) above
for the Gaussian initial distribution.

In the case of the attractive nonlinearity, $\sigma =-1$, with the same set
of parameters as above, the initial real wave function (\ref{CombGauss}) was
quickly, within $t\approx 5$, transformed into a complex one, with a
subsequent quasi-random energy exchange between the real and imaginary
parts. At the initial stage of the evolution, the central part of the
solution would transform into a very narrow large-amplitude pulse, with two
side wings represented by small-amplitude wide pulses. Then, the central
pulse would oscillate in time quasi-randomly.

\section{Exact solutions for the two-dimensional stationary
Gross--Pitaevskii equation}
\label{sect3}

In this section we proceed to the consideration of the 2D version of the
stationary GPE, taken in the dimensionless form similar to Eq.~(\ref{St1DGPE}%
), with spatial coordinates $\xi $ and $\eta $:
\begin{equation}
\varphi _{\xi \xi }+\varphi _{\eta \eta }=\left[ u(\xi ,\eta )-\mu \right]
\varphi +\sigma |\varphi |\,^{2}\varphi .  \label{St2DGPE}
\end{equation}%
The approach similar to that which was developed in subsection \ref{ss2.E}
can be employed to construct an appropriate 2D potential on the basis of a
given solution. From Eq.~(\ref{St2DGPE}) one formally deduces
\begin{equation}
u(\xi ,\eta )=\left( \varphi _{\xi \xi }+\varphi _{\eta \eta
}\right)/\varphi -\sigma \varphi \,^{2}+\mu .  \label{2dPotLump}
\end{equation}%
Examples presented below aim to demonstrate solutions which may be useful to
physical applications.

\begin{figure}[t!]
\vspace*{-25mm}
\centerline{\psfig{figure=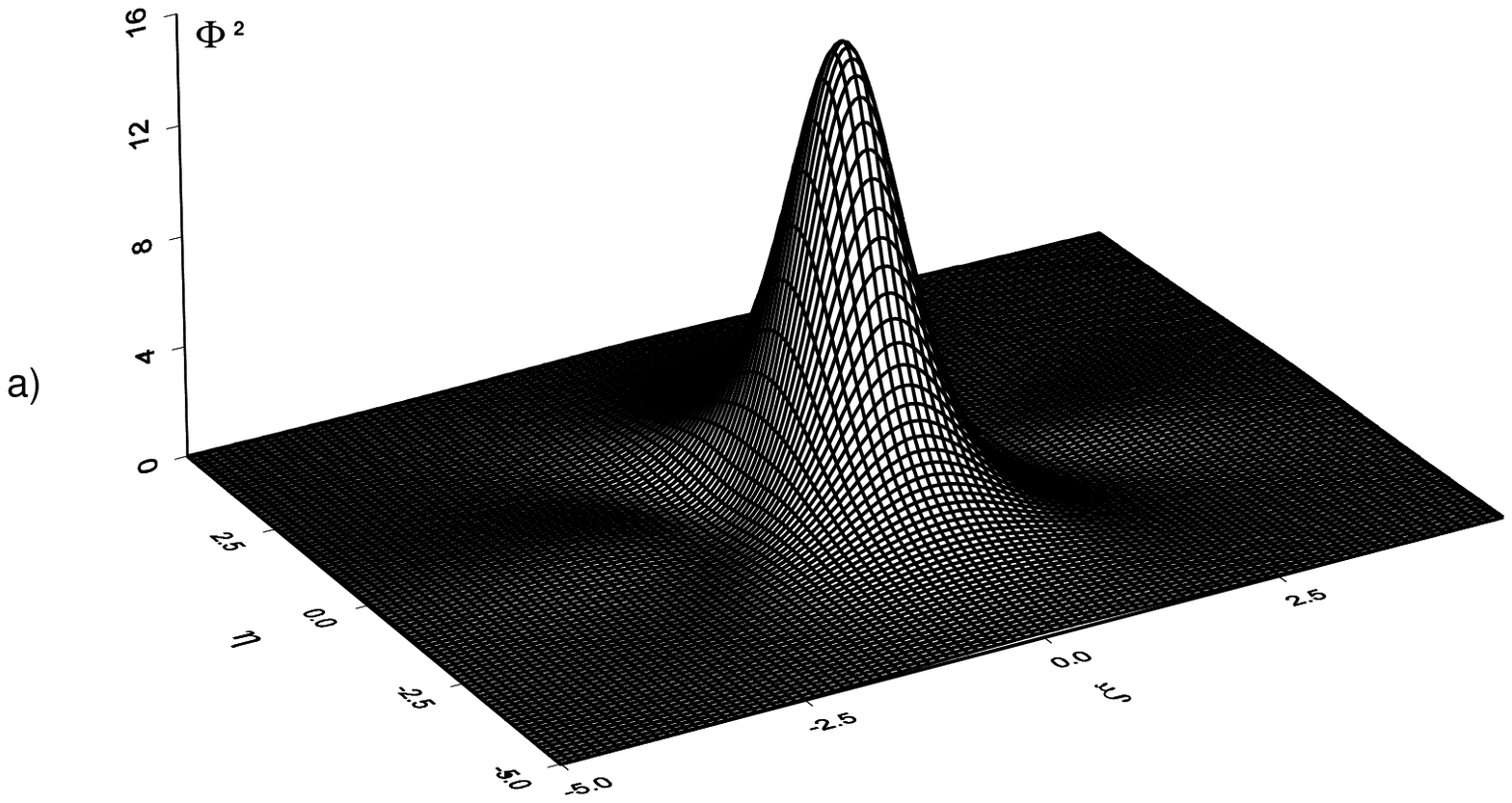,height=70mm}} %
\vspace{5mm}%
\centerline{\psfig{figure=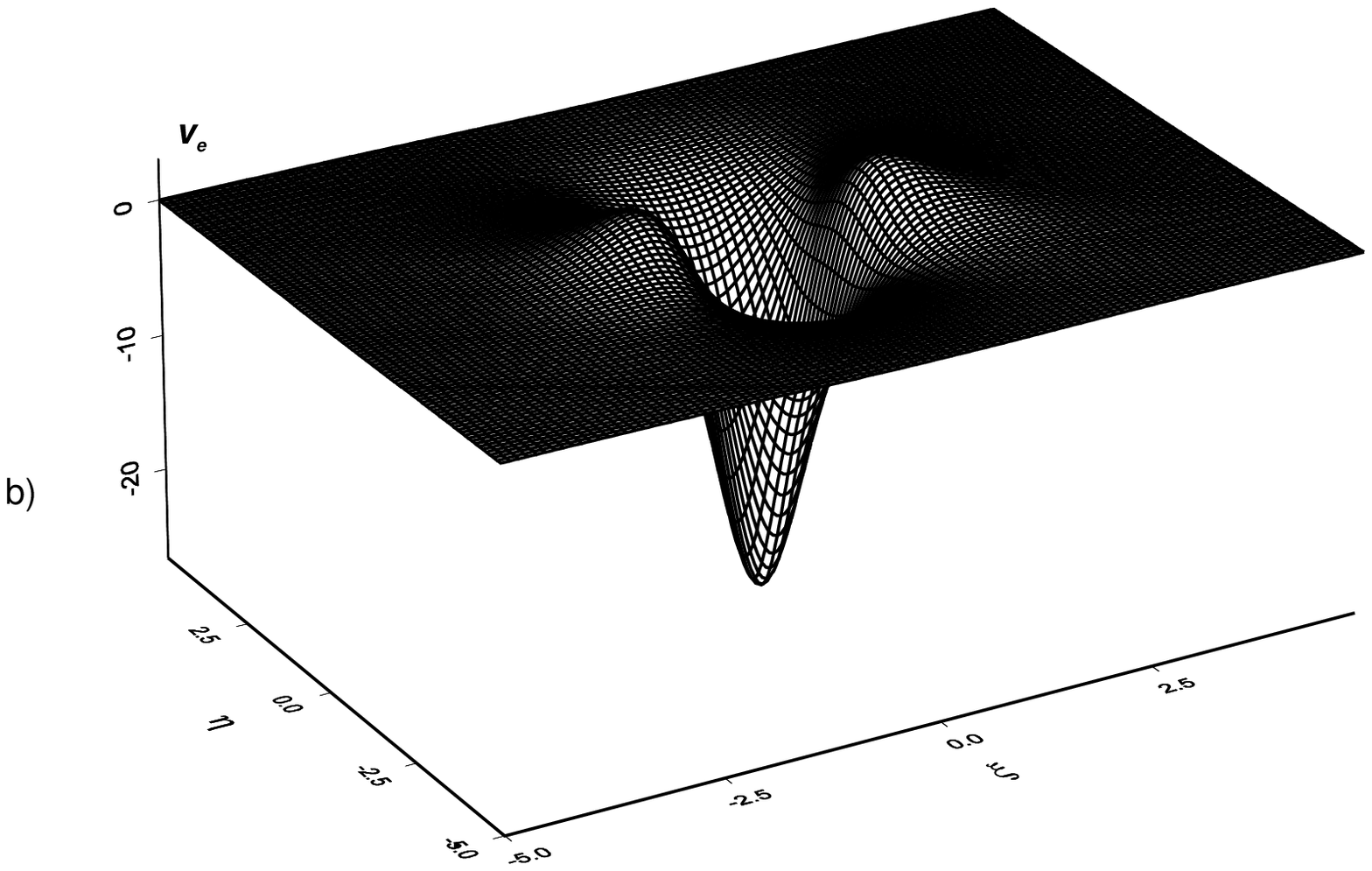,height=70mm}}%
\vspace{8mm}%
\centerline{\psfig{figure=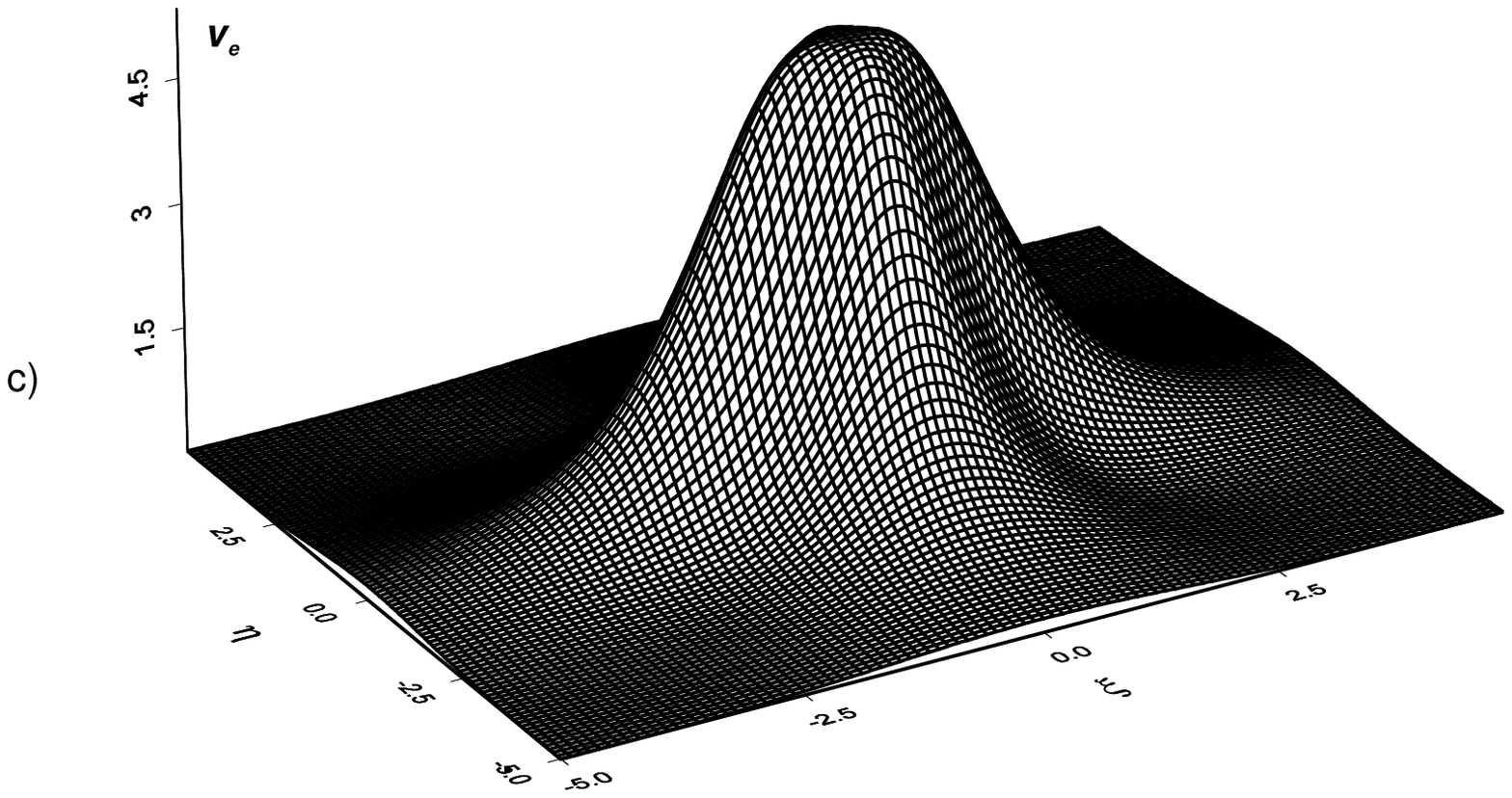,height=70mm}}%
\caption{(a) Lump solution (\protect\ref{Lump})
shown in terms of $\left[ \Phi \left( \protect\xi ,\protect\eta
\right) /A\right]
^{2} $ for $a=b$ \ and $M=0$; and the corresponding potential (\protect\ref%
{2dNSLumpPot}) for $\protect\sigma =1$ (b) and $\protect\sigma = -1$ (c).}%
\label{f14}
\end{figure}

\emph{The Kadomtsev--Petviashvili lump soliton.} As the first trial
function, we take an anisotropic 2D weakly localized ansatz known as the
\textit{lump }solution to the KP1 equation \cite{AblSegur}:
\begin{equation}
\Phi (\xi ,\eta )=12A\frac{3-(\xi /a)^{2}+(\eta /b)^{2}}{\left[ 3+(\xi
/a)^{2}+(\eta /b)^{2}\right] ^{2}}.  \label{Lump}
\end{equation}%
Substituting it as $\varphi $ into Eq.~(\ref{2dPotLump}), one obtains the
corresponding potential,
\begin{equation}
v_{e}(\bar{\xi},\bar{\eta})=-\frac{6}{S^{2}}\frac{P_{4}(\bar{\xi},\bar{\eta})%
}{(3+\bar{\xi}^{2}+\bar{\eta}^{2})^{2}(3-\bar{\xi}^{2}+\bar{\eta}^{2})}%
-144\sigma \frac{\left( 3-\bar{\xi}^{2}+\bar{\eta}^{2}\right) ^{2}}{\left( 3+%
\bar{\xi}^{2}+\bar{\eta}^{2}\right) ^{4}}+M,  \label{DimLump}
\end{equation}
where $v_{e}=u/A^{2}$, $\bar{\xi}=\xi /a$, $\bar{\eta}=\eta /b$, $S=Ab$, $%
\beta =b/a$, $M=\mu /A^{2}$, and $P_{4}(\bar{\xi},\bar{\eta})=(\beta ^{2}-1)(%
\bar{\xi}^{4}+\bar{\eta}^{4})+6(1-\beta ^{2})(\bar{\xi}\bar{\eta})^{2}-2(%
\bar{\xi}^{2}+\bar{\eta}^{2})-9\beta ^{2}(2\bar{\xi}^{2}+\bar{\eta}%
^{2})+3(1+3\beta ^{2})$. In the particular case of $\beta =1$ ($a=b$), Eq.~(%
\ref{DimLump}) simplifies to
\begin{equation}
v_{e}(\bar{\xi},\bar{\eta})=-144\frac{2(3-5\bar{\xi}^{2}+\bar{\eta}%
^{2})+\sigma (3-\bar{\xi}^{2}+\bar{\eta}^{2})^{2}}{(3+\bar{\xi}^{2}+\bar{\eta%
}^{2})^{4}}+M.  \label{2dNSLumpPot}
\end{equation}%
Both expressions (\ref{DimLump}) and (\ref{2dNSLumpPot}) correspond to
anisotropic 2D trapping potentials. Figure \ref{f14}a shows a 3D view of
lump solution (\ref{Lump}) for $a=b$, and Figs.~\ref{f14}b,c display the
corresponding potential (\ref{2dNSLumpPot}) for $\sigma =\pm 1$. In the case $\sigma =-1$,
the potential represents a 2D hump, i.e. it is repulsive,
hence the corresponding solution (\ref{Lump}) is apparently unstable.

\emph{2D Gaussian trial function.} Another natural example in the 2D case is
provided by the solution ansatz in the form of an axisymmetric Gaussian, $%
\Phi (r)=A\exp {(-r^{2}/l^{2})}$, where $r^{2}=\xi ^{2}+\eta ^{2}$. From
Eq.~(\ref{St2DGPE}) we deduce the potential in the dimensionless form:
\begin{equation}
v_{e}(\rho ) = \frac{4}{S^{2}}\left( \rho ^{2}-1\right) -\sigma \exp {%
(-2\rho ^{2})}+M,  \label{2DGauss}
\end{equation}%
where $v_{e}=u/A^{2}$, $\rho =r/l$, $S=Al$ and $M=\mu /A^{2}$ (the same
particular solution was recently obtained in Ref.~\cite{WangHao09} in a
different way, and its stability has been established in direct simulation).
The principal cross-sections of the squared Gaussian solution and the
corresponding potentials for $\sigma =\pm 1$ are shown in Fig.~\ref{f15} for
three values of $S$. In the case of the attractive nonlinearity, $\sigma =-1$%
, potential function $v_{e}$ may feature a local maximum at the center,
which appears at $S>\sqrt{2}$ (see Fig.~\ref{f15}c).

\clearpage

\begin{figure}[t]%
\vspace*{-30mm}%
\begin{tabular}{cc}
\hspace*{-15mm}\psfig{figure=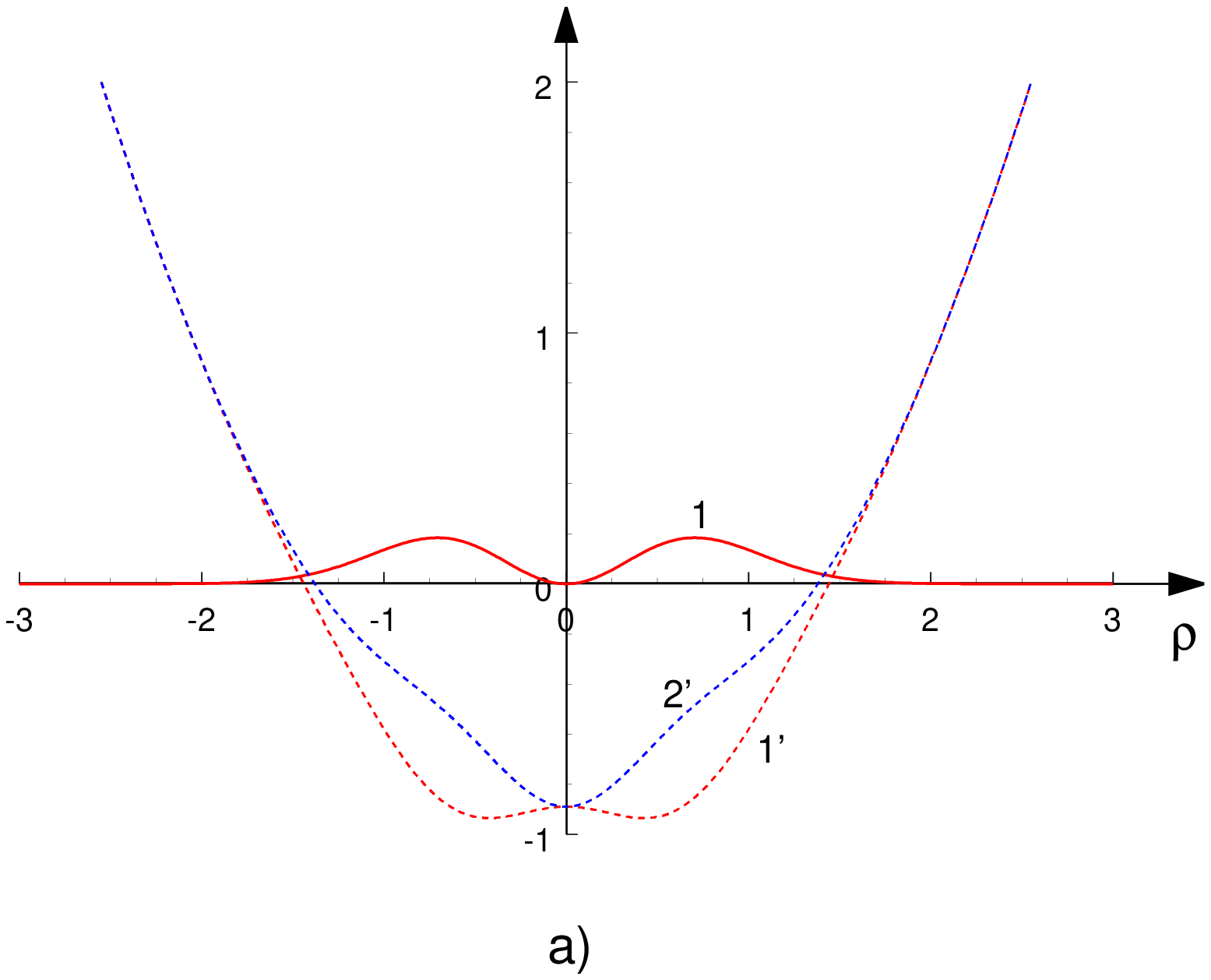,height=65mm} & \psfig{figure=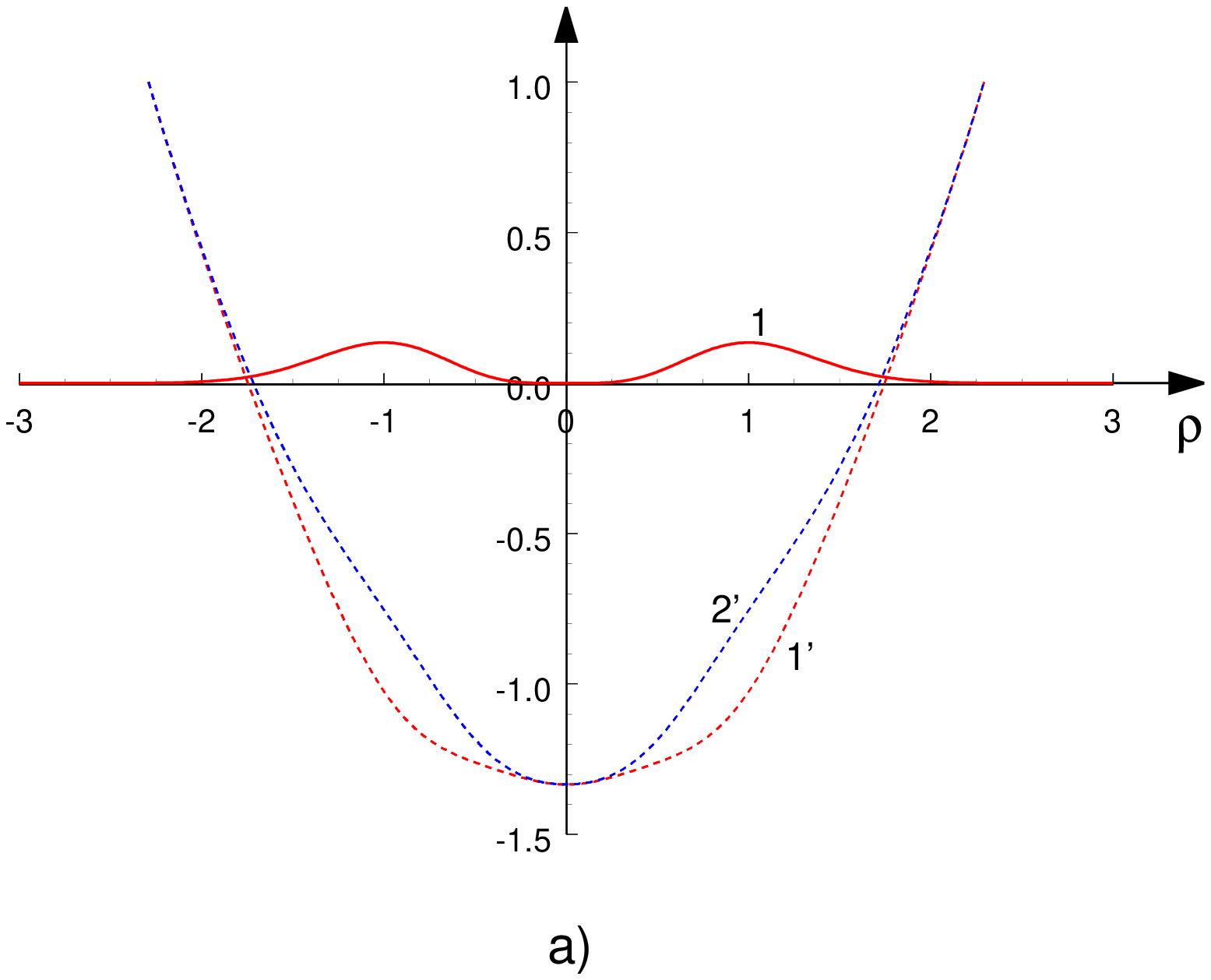,height=65mm}
\\
\hspace*{-15mm}\psfig{figure=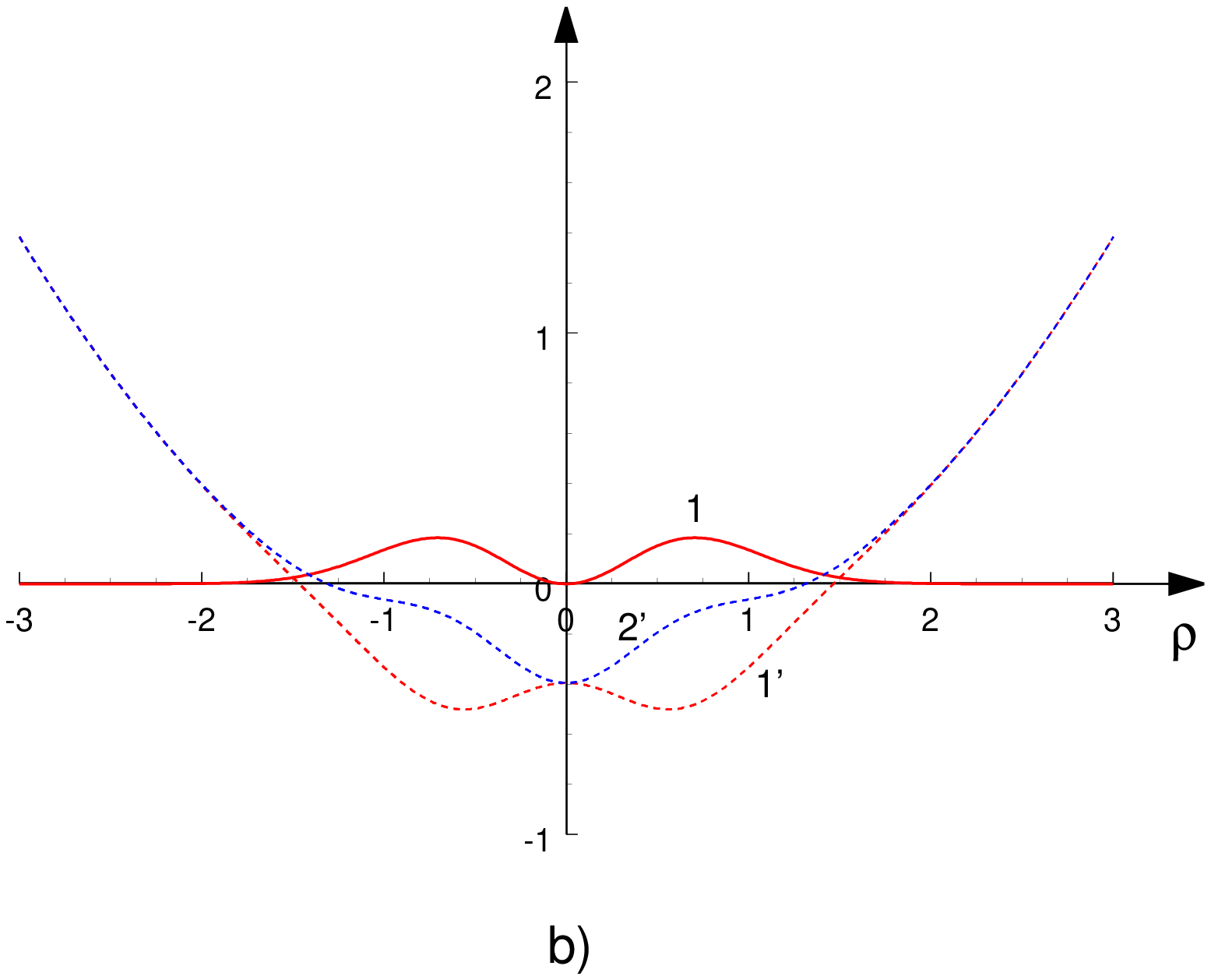,height=65mm} & \psfig{figure=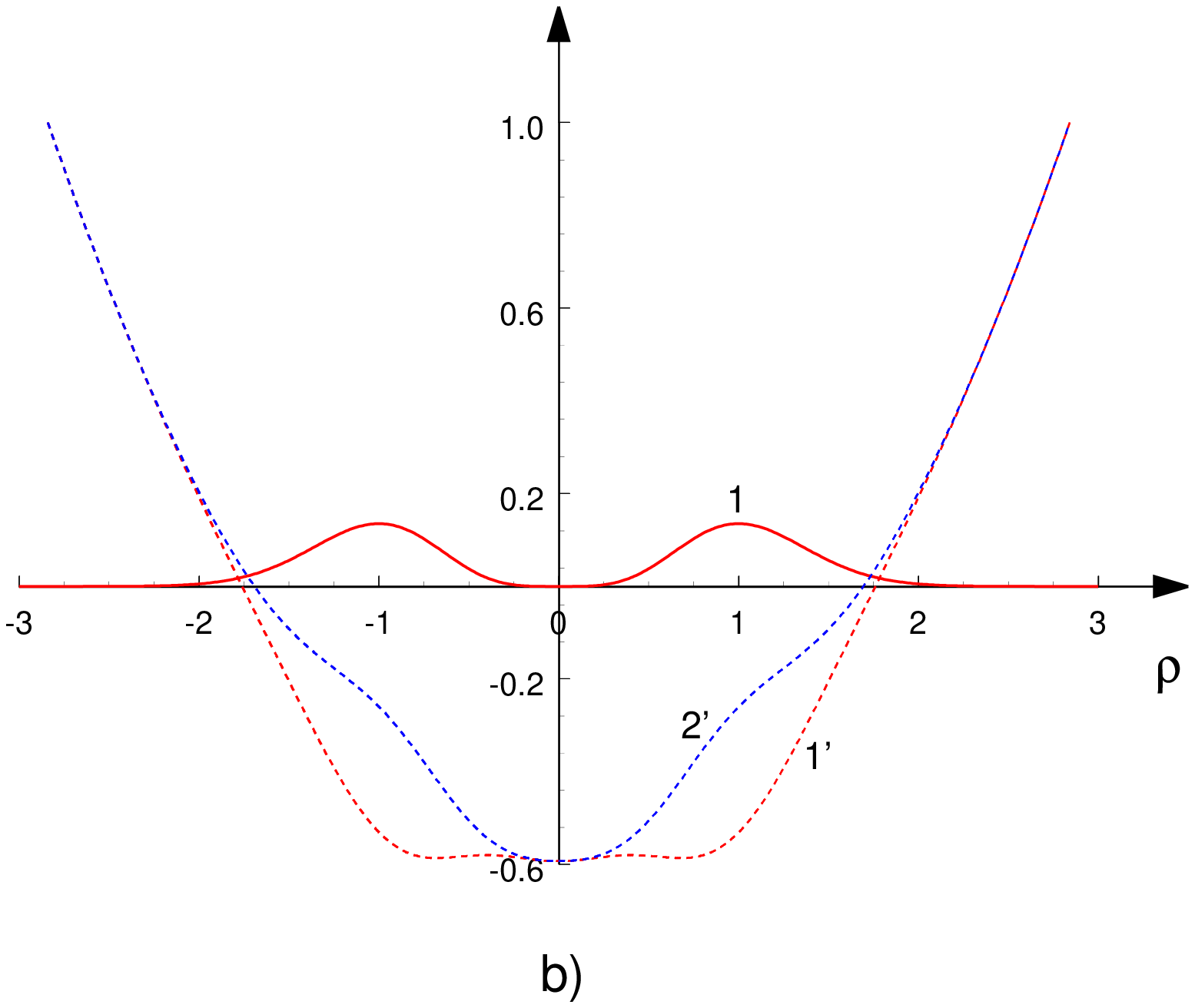,height=65mm}
\\
\hspace*{-15mm}\psfig{figure=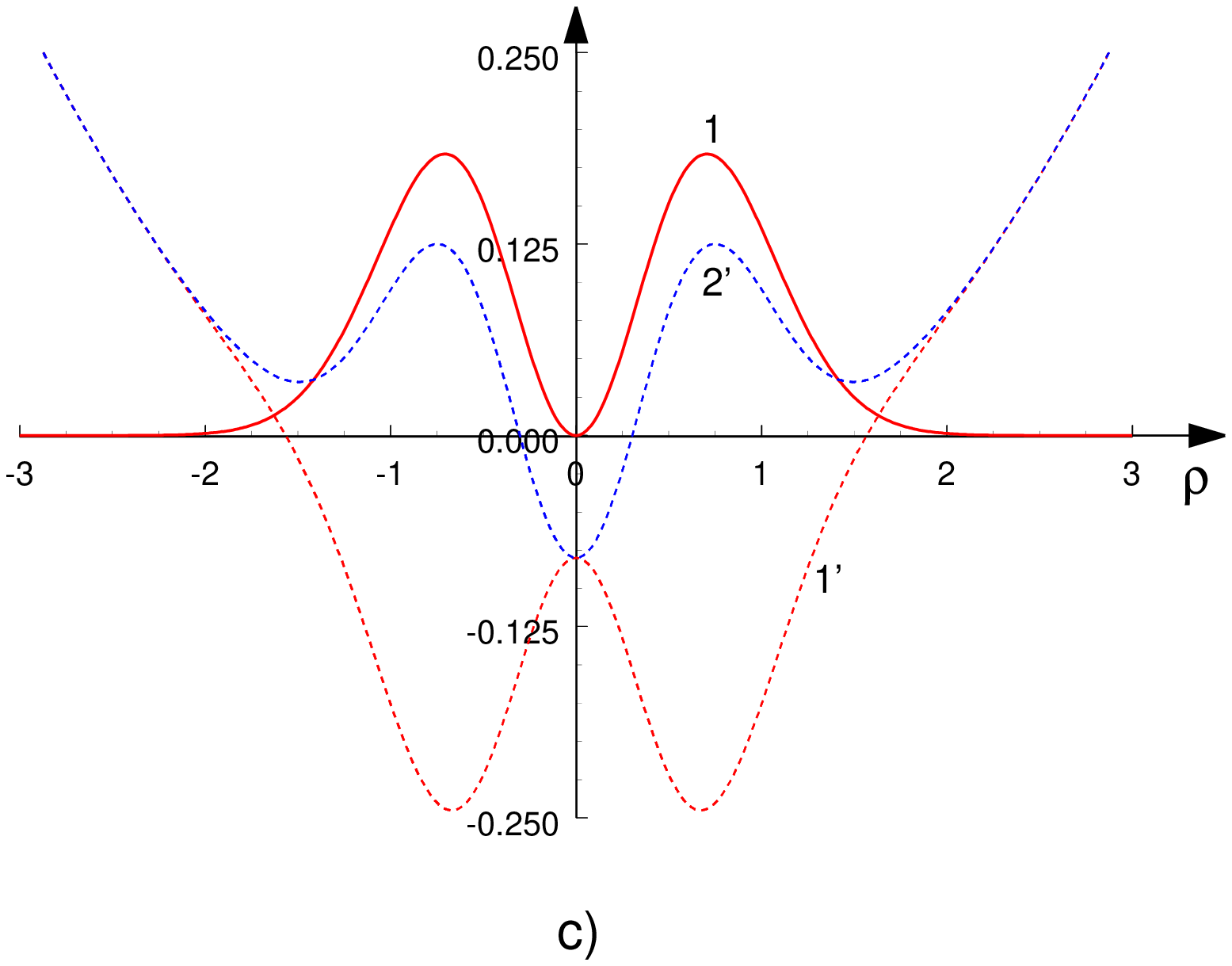,height=65mm} & \psfig{figure=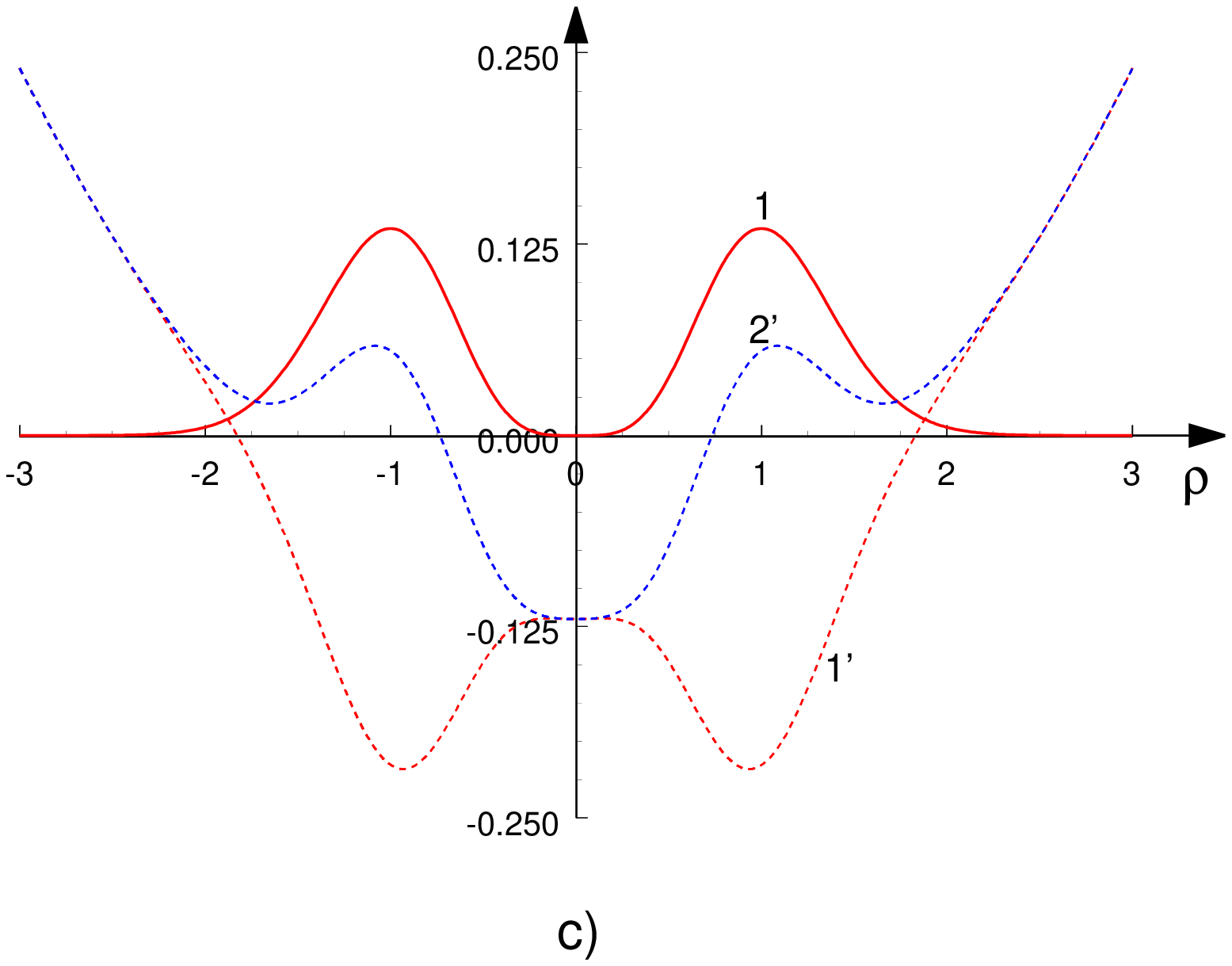,height=65mm}
\\
FIG.~17 & FIG.~18%
\end{tabular}%
\caption{The radial distribution of density $|\protect\varphi %
(\protect\rho )|^{2}$ in the unitary vortex solution (lines $1$),
and the corresponding potentials (\protect\ref{ExplRealPot}) with
$M=0$ for the
repulsive ($\protect\sigma =1$, lines $1^{\prime }$) and attractive ($%
\protect\sigma =-1$, lines $2^{\prime }$) nonlinearity. a) $S=3$, b) $S=4.5$%
, c) $S=10$.}
\label{f16}%
\caption{The radial distribution of density $|\protect\varphi %
(\protect\rho )|^{2}$ for the double vortex (lines $1$), and the
corresponding potentials (\protect\ref{2ChargePot}) with $M=0$ for the
repulsive ($\protect\sigma =1$, lines $1^{\prime }$) and attractive ($%
\protect\sigma =-1$, lines $2^{\prime }$) nonlinearities. a) $S=3$, b) $%
S=4.5 $, c) $S=10$.}
\label{f17}
\end{figure}

Other examples represent 2D patterns of a different type, namely, \emph{%
vortices}. First we present the unitary vortex, with topological charge $J=1$
(it may be treated as a 2D counterpart of the 1D derivative-Gaussian profile
considered above): $\Phi (\xi ,\eta )=A(\xi +i\eta )\exp {(-r^{2}/l^{2})}$,
where $r=\sqrt{\xi ^{2}+\eta ^{2}}$. For the complex solution it is
convenient to cast the stationary GPE (\ref{St2DGPE}) into the following
real form: $(1/2)\left[ \Delta \left( |\varphi |^{2}\right) -2|\nabla
\varphi |^{2}\right] -\sigma |\varphi |^{4}=[u(\xi ,\eta )-\mu ]|\varphi
|^{2},$ where $\Delta $ is the 2D Laplacian. From this equation, one can
deduce the potential:
\begin{equation}
u(\xi ,\eta ) = \frac{\Delta |\varphi |^{2}-2|\nabla \varphi |^{2}}{
2|\varphi |^{2}} -\sigma |\varphi |^{2}+\mu .  \label{RealPot}
\end{equation}%
Substituting here the wave function of the vortex, the potential can be
obtained in the explicit form [cf. Eq.~(\ref{InvPotS2})]:
\begin{equation}
v_{e}(\rho )=\left( 4/S^{2}\right) \left( \rho ^{2}-2\right) -\sigma \rho
^{2}\exp {(-2\rho ^{2})}+M,  \label{ExplRealPot}
\end{equation}%
where $v_{e}=u/(Al)^{2}$, $\rho =r/l$, $S=Al^{2}$, and $M=\mu /(Al)^{2}$.
Principal cross-sections of the solution for $|\Phi (\rho )|^{2}$ and the
corresponding potentials for $\sigma =\pm 1$ are shown in Fig.~\ref{f16},
for three values of $S$.

\textit{The double vortex}. We have also considered the trial
solution in the form of the vortex with $J=2$, \textit{viz}.,
$\Phi (\xi ,\eta
)=A(x+iy)^{2}\exp {(-r^{2}/l^{2})}$. One can readily deduce from Eq.~(\ref%
{RealPot}) the potential supporting this solution:
\begin{equation}
v_{e}(\rho )=\left( 4/S^{2}\right) \left( \rho ^{2}-3\right)
-\sigma \rho ^{4}\exp {(-2\rho ^{2})}+M,  \label{2ChargePot}
\end{equation}
where $v_{e}=u/(Al^{2})^{2}$, $\rho =r/l$, $S=Al^{3}$, and $M=\mu
/(Al^{2})^{2}$. Principal cross-sections of the solution for
$|\Phi (\rho )|^{2}$ and the corresponding potentials for $\sigma
=\pm 1$ are shown in Fig.~\ref{f17} for the same three values of
$S$ as in Fig.~\ref{f16}.

\section{Exact solutions for the three-dimensional stationary
Gross--Pitaevskii equation}
\label{sect5}

In this section we briefly demonstrate the applicability of the above method
to the 3D case (with the usual Laplacian corresponding to the GPE). The
scaled 3D version of the stationary GPE is

\begin{equation}
\frac{\partial \,^{2}\varphi }{\partial \xi \,^{2}}+\frac{\partial
\,^{2}\varphi }{\partial \eta \,^{2}}+\frac{\partial \,^{2}\varphi }{%
\partial \zeta \,^{2}}=\left[ u(\xi ,\eta )+\mu \right] \varphi +\sigma
|\varphi |\,^{2}\varphi ,  \label{St3DGPE}
\end{equation}%
where function $\varphi $ depends on three spatial coordinates -- $\xi $, $%
\eta$, and $\zeta $. From Eq.~(\ref{St3DGPE}) one can readily deduce the
necessary potential:

\begin{figure}[t]
\vspace*{-20mm}%
\begin{tabular}{c}
\hspace*{40mm}\psfig{figure=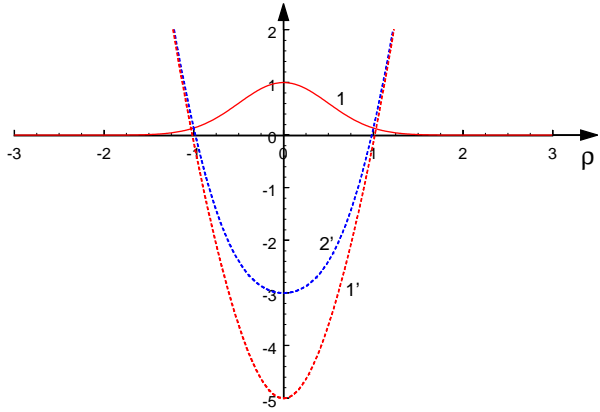,height=65mm} \\ %
\hspace*{40mm}\psfig{figure=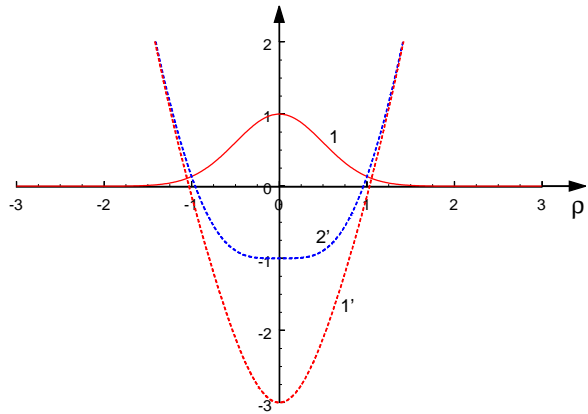,height=65mm} \\ %
\hspace*{40mm}\psfig{figure=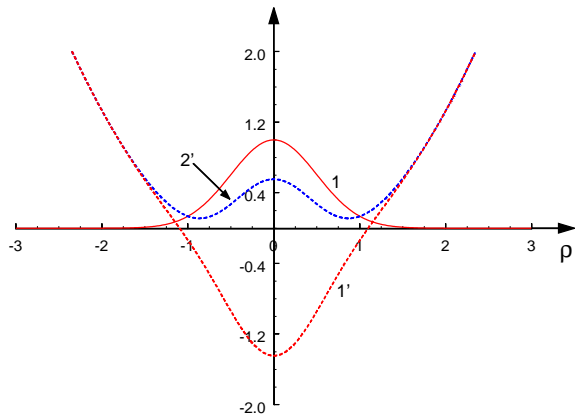,height=65mm}%
\end{tabular}%
\caption{The radial distribution of density $|\protect\varphi %
(\protect\rho )|^{2}$ for the 2D Gaussian solution (lines $1$ in each panel)
and the corresponding potentials (\protect\ref{2DGauss}) with $M=0$ for the
repulsive ($\protect\sigma =1$, lines $1^{\prime }$) and attractive ($%
\protect\sigma =-1$, lines $2^{\prime }$) nonlinearities: a) $S=1$, b) $S=%
\protect\sqrt{2}$, c) $S=3$.}
\label{f15}
\end{figure}

\clearpage

\clearpage

\begin{equation}  \label{2dPotLump2}
u(\xi,\eta) = \frac{1}{\varphi}\left(\frac{\partial\,^2\varphi}{%
\partial\xi\,^2} + \frac{\partial\,^2\varphi}{\partial\eta\,^2}+ \frac{%
\partial\,^2\varphi}{\partial\zeta\,^2}\right) - \sigma\varphi\,^2 - \mu.
\end{equation}

We here present an example of the 3D trial function in the form of the
spherically symmetric Gaussian, $\Phi (r)=A\exp {(-r^{2}/l^{2})}$, where $%
r^{2}=\xi ^{2}+\eta ^{2}+\zeta ^{2}$. The corresponding potential can be
found from Eq.~(\ref{St3DGPE}):

\begin{equation}
v_{e}(\rho )=\frac{4}{S^{2}}\left( \rho ^{2}-\frac{3}{2}\right) -\sigma \exp
{(-2\rho ^{2})}-M,  \label{3DGauss}
\end{equation}%
where $v_{e}=u/A^{2}$, $\rho =r/l$, $S=Al$, and $M=\mu /A^{2}$. The plot of
the potential, $v_{e}(\rho)$, is similar to that shown in Fig.~\ref{f15}. In
a similar way, one can readily construct potentials for many other cases
including 3D generalizations of the examples presented above in terms of the
2D geometry.

\section{Conclusions}
\label{sect6}

We have demonstrated that numerous exact 1D stationary solutions
to the GPE (Gross--Pitaevskii equation) may be constructed with
the help of the Kondrat'iev--Miller method \cite{KondrMiller}.
Within the framework of this method, the corresponding potential
function in the GPE is proportional to stationary solution
$\varphi (\xi )$, which, in turn, was taken as a solution to the
stationary GE (Gardner equation). The stability of the 1D
solutions was tested through direct simulations of the
time-dependent GPE. It was found that some solitary-type solutions
are stable -- in particular, those corresponding to the solution
of the GE in the form of the ``fat" (table-top) soliton, which is
given by Eq.~(\ref{FatSol}), in the case of the repulsive
nonlinearity. A stable soliton solution in the case of the
attractive nonlinearity was also found, \textit{viz}., the
exponentially localized solution given by expression
(\ref{ExpSol}).

Further, we have proposed an ``inverse method" for the GPE, as a
way to construct appropriate potentials for a given distribution
of the wave function. It was demonstrated that this method helps
to produce many solutions in 1D and 2D settings. The stability of
all the so found 1D solutions has been tested in direct
simulations. The 1D and 2D potentials constructed here as supports
for basic natural types of the localized matter-wave distributions
are fairly simple, and may be realized in the experiment by means
of currently available techniques, based on the design of
appropriate magnetic and optical traps for the BEC.

The numerical scheme employed here for the simulations of the time-dependent
GPE in 1D is based on the Yunakovsky's method of the operator exponential,
which has been used in many previous works (see, e.g., Ref. \cite{Yunakovsky}%
). The method is also efficient in obtaining solutions to NLSE and GPE in
the space of any dimension. It is briefly described in Appendix.

In addition to testing the stability of the 2D localized solutions, another
remaining issue is to check whether the specially designed potentials which
support 1D and 2D solutions, as ground states, may also sustain higher-order
bound states in the respective nonlinear Further, the inverse method can be
readily extended to the 3D settings. Some results have been already obtained
in this direction, to be reported elsewhere. Finally, it may be quite
interesting to apply both the GE and the inverse method to constructing
exact solutions for dark solitons in 1D and circular dark solitons in 2D
\cite{dark}, in the case of the modulationally stable background. These
generalizations will be also reported elsewhere. \newline

\textbf{Acknowledgements.} This work was partially supported by the
German--Israel Foundation through the grant No. 149/2006. Y.S. appreciates
hospitality of the Faculty of Engineering at the Tel Aviv University during
his visit in 2009.

\section{Appendix: The numerical algorithm for simulations of the
Gross-Pitaevskii equation}

In this Appendix we describe a numerical algorithm based on the method
originally developed by Yunakovsky for the numerical solution of the NLSE
\cite{Yunakovsky}. The method works equally well in 1D, 2D and 3D settings.
Below give a brief account of the method in the application to the 1D case,
it generalization for 2D and 3D settings being straightforward.

One starts by the application of the Fourier transform to variable $\xi $ in
Eq.~(\ref{dl1DGPE}):
\begin{equation}
i\frac{\partial \widetilde{\varphi }}{\partial t}=k^{2}\widetilde{\varphi }+%
\hat{F}\left\{ [u(\xi )+\mu ]\varphi \right\} +\sigma \hat{F}\left\{
|\varphi |^{2}\varphi \right\} ,  \label{FGPE}
\end{equation}%
where $k$ is the respective wavenumber, and the tilde stands for the Fourier
image, which is generated by the Fourier-transform operator $\hat{F}$. Next,
we introduce a new function, $u(k,t)=\widetilde{\varphi }(k,t)\exp {(ik^{2}t)%
}$, and rewrite Eq.~(\ref{FGPE}) accordingly:
\begin{equation}
i\frac{\partial u}{\partial t}=\hat{F}\left\{ \left[ U_{\mu }(\xi )+\sigma
|\varphi |^{2}\right] \varphi \right\} \exp {(ik^{2}t)},  \label{FGPEu}
\end{equation}%
where $U_{\mu }(\xi )\equiv u(\xi )+\mu $. This equation may be formally
integrated in $t$, yielding, in terms of $\varphi $,
\begin{equation}
\widetilde{\varphi }(k,t)=\widetilde{\varphi }(k,0)\exp {(-ik^{2}t)}%
-i\int\limits_{0}^{t}\hat{F}\left\{ \left[ U_{\mu }(\xi )+\sigma |\varphi |^{2}%
\right] \varphi \right\} \exp {[-ik^{2}(t-\tau )]}\,d\tau .  \label{FGPEfIn}
\end{equation}%
The integral on the right-hand side can be approximately calculated, over a
small time interval, with the help of the trapezoidal rule. The result,
valid up to $O(t^{3})$, is
\begin{gather}
\widetilde{\varphi }(k,t)=\widetilde{\varphi }_{0}\exp {(-ik^{2}t)}%
-(it/2){}{}  \notag \\
\times \left[ \hat{F}\left\{ \left[ U_{\mu }(\xi )+\sigma |\varphi _{0}|^{2}%
\right] \varphi _{0}\right\} \exp {(-ik^{2}t)}+\hat{F}\left\{ \left[ U_{\mu
}(\xi )+\sigma |\varphi |^{2}\right] \varphi \right\} \right] .
\label{Trapez}
\end{gather}%
Next, we collect on the left-hand side those terms which depend on the
current time, and leave on the right-hand side the terms which depend on
initial conditions:
\begin{eqnarray}
\widetilde{\varphi }(k,t)+\frac{it}{2}\hat{F}\left\{ \left[ U_{\mu }(\xi
)+\sigma |\varphi |^{2}\right] \varphi \right\} = &&  \notag \\
\left( \widetilde{\varphi }_{0}-\frac{it}{2}\hat{F}\left\{ \left[ U_{\mu
}(\xi )+\sigma |\varphi _{0}|^{2}\right] \varphi _{0}\right\} \right) \exp {%
(-ik^{2}t)}. &&{}  \label{Collection}
\end{eqnarray}%
By applying the inverse Fourier transform, $\hat{F}^{-1}$, to Eq.~(\ref%
{Collection}), one obtains
\begin{eqnarray}
&&\left\{ 1+\frac{it}{2}\left[ U_{\mu }(\xi )+\sigma |\varphi (\xi ,t)|^{2}%
\right] \right\} \varphi (\xi ,t)  \notag \\
&=&\hat{F}^{-1}\left\{ \left[ \widetilde{\varphi }_{0}(k)-\frac{it}{2}\hat{F}%
\left\{ \left[ U_{\mu }(\xi )+\sigma |\varphi _{0}|^{2}\right] \varphi
_{0}\right\} \right] \exp {(-ik^{2}t)}\right\} \equiv B(\xi ,t).
\label{InFTrap}
\end{eqnarray}%
Function $B(\xi ,t)$ produced by Eq.~(\ref{InFTrap}) is an explicit result
obtained from the given initial condition, $\varphi _{0}(k)\equiv \widetilde{%
\varphi }(\xi ,0)$. Then, function $\varphi (\xi ,t)$ can be formally found
from Eq.~(\ref{InFTrap}):
\begin{equation}
\varphi (\xi ,t)=\frac{B(\xi ,t)}{1+(it/2)\left[ U_{\mu }(\xi )+\sigma
|\varphi (\xi ,t)|^{2}\right] }.  \label{FunctPhi}
\end{equation}%
To make this formula practical, one needs to define $|\varphi (\xi ,t)|^{2}$
in the denominator of Eq.~(\ref{FunctPhi}). To do that, take Eq.~(\ref%
{InFTrap}) and multiply it by the complex conjugate counterpart, which
yields
\begin{equation}
\left\{ 1+\left( t^{2}/4\right) \left[ U_{\mu }(\xi )+\sigma |\varphi |^{2}%
\right] ^{2}\right\} |\varphi (\xi ,t)|^{2}=|B(\xi ,t)|^{2}.  \label{ModIFT}
\end{equation}%
Thus, denoting $z\equiv |\varphi (\xi ,t)|^{2}$ and taking into account that
$\sigma ^{2}=1$, we obtain a cubic equation for $z$:
\begin{equation}
z^{3}+2\sigma U_{\mu }(\xi )z^{2}+\left[ \left( 4/t^{2}\right) +U_{\mu }(\xi
)\right] z-\left( 4/t^{2}\right) |B(\xi ,t)|^{2}=0.  \label{CubEq}
\end{equation}%
The cubic equation can be solved analytically, in principle. Its single real
root (two others are complex) can be found by means of symbolic calculations
realized by means of software such as Maple:%
\begin{equation}
z=\frac{2}{3}\left[ \mathrm{Root3}-\frac{1/t^{2}+U_{\mu }(\xi )/4-U_{\mu
}^{2}(\xi )/3}{\mathrm{Root3}}-\sigma U_{\mu }(\xi )\right] ,  \label{RealRt}
\end{equation}%
where we define
\begin{eqnarray}
\mathrm{Root3} &=&\left\{ \frac{9}{8}\sigma U_{\mu }(\xi )\left[ \frac{4}{%
t^{2}}+U_{\mu }(\xi )\right] +\frac{27}{4t^{2}}|B(\xi ,t)|^{2}-\sigma U_{\mu
}^{3}(\xi )+\frac{3}{8t^{3}}\sqrt{X}\right\} ^{1/3},  \notag \\
&& \nonumber \\ %
X &=&3\left\{ \left[ 1-U_{\mu }(\xi )\right]
U_{\mu }^{3}(\xi )t^{6}+4\left[
3-2U_{\mu }(\xi )+\sigma \left( 9-8U_{\mu }(\xi )\right) |B(\xi ,t)|^{2}%
\right] U_{\mu }^{2}(\xi )t^{4}+{}\right.  \notag \\
{} &&\left. \left[ 48U_{\mu }(\xi )+\left( 144\sigma U_{\mu }(\xi
)+108\right) |B(\xi ,t)|^{2}-168\sigma U_{\mu }^{2}(\xi )\right]
t^{2}+64\right\} .  \label{SqRt}
\end{eqnarray}%
Once root $z$ was found, it can be substituted into the denominator of Eq.~(%
\ref{FunctPhi}); then, function $\varphi (\xi ,t)$ is completely determined
at time $t$. After that, the procedure may be repeated for the next time
step.

%\clearpage


\begin{thebibliography}{99}
\bibitem{PitString03} L.~P.~Pitaevskii and S.~Stringari, \textit{%
Bose--Einstein Condensation}. (Clarendon, Oxford, 2003).

\bibitem{finite} M. M. Cerimele, M. L. Chiofalo, F. Pistella, S. Succi, and
M. P. Tosi, Phys. Rev. E \textbf{62}, 1382 (2000).

\bibitem{split-step} S. K. Adhikari and P. Muruganandam, J. Phys. B: At.
Mol. Opt. Phys. \textbf{35}, 2831 (2002); P. Muruganandam and S. K.
Adhikari, \textit{ibid}. \textbf{36}, 2501 (2003); W. Z. Bao, D. Jaksch, and
P. A. Markowich, J. Comp. Phys. \textbf{187}, (2003).

\bibitem{Crank} S. K. Adhikari, Phys. Rev. E \textbf{65}, 016703 (2002).

\bibitem{Chiofalo00} M.~L.~Chiofalo, S.~Succi, and M.~P.~Tosi,
%Ground state
%of trapped interacting Bose-Einstein condensates by an explicit
%imaginary-time algorithm,
Phys. Rev. E \textbf{62}, 7438 (2000); W. Z. Bao and Q. Du, SIAM J. Scient.
Comp. \textbf{25}, 1674 (2004).

\bibitem{Minguzzi04} A.~Minguzzi, S.~Succi, F.~Toschi, M.~P.~Tosi, and
P.~Vignolo, Phys. Rep. \textbf{395}, 223 (2004).

\bibitem{VA} V. M. P\'{e}rez-Garc\'{\i}a, H. Michinel, J. I. Cirac, M.
Lewenstein, and P. Zoller, Phys. Rev. A \textbf{56}, 1424 (1997); A. L.
Fetter, J. Low Temp. Phys. \textbf{106}, 643 (1997); B. A. Malomed, in
Progr. Optics, ed. by E. Wolf, vol. 43, p. 71 (North Holland, Amsterdam,
2002); S. K. Adhikari and B. A. Malomed, Phys. Rev. A \textbf{77}, 023607
(2008).

\bibitem{Luca} L. Salasnich, A. Parola, and L. Reatto, Phys. Rev. A \textbf{%
65}, 043614 (2002); L. Salasnich and B. A. Malomed, \textit{ibid}. \textbf{74%
}, 053610 (2006); A. Mu\~{n}oz Mateo and V. Delgado, \textit{ibid}. \textbf{%
77}, 013617 (2008).

\bibitem{discrete} A. Trombettoni and A. Smerzi, Phys. Rev. Lett. \textbf{86}%
, 2353 (2001); F. Kh. Abdullaev, B. B. Baizakov, S. A. Darmanyan, V. V.
Konotop, and M. Salerno, Phys. Rev. A \textbf{64}, 043606 (2001); G. L.
Alfimov, P. G. Kevrekidis, V. V. Konotop, and M. Salerno, Phys. Rev. E
\textbf{66}, 046608 (2002); R. Carretero-Gonz\'{a}lez and K. Promislow,
Phys. Rev. A \textbf{66}, 033610 (2002); M. A. Porter, R. Carretero-Gonz\'{a}%
lez, P. G. Kevrekidis, and B. A. Malomed, Chaos \textbf{15}, 015115 (2005);
A. Maluckov, L. Had\v{z}ievski, B. A. Malomed, and L. Salasnich, Phys. Rev.
A \textbf{78}, 013616 (2008).

\bibitem{TF} F. Dalfovo, C. Minniti, S. Stringari, L. Pitaevskii, Phys.
Lett. A \textbf{227}, 259 (1997).

\bibitem{DWP} E. A. Ostrovskaya, Y. S. Kivshar, M. Lisak, B. Hall, F.
Cattani, and D. Anderson, Phys. Rev. A \textbf{61}, 031601(2000); R.
D'Agosta, B. A. Malomed, C. Presilla, Phys. Lett. A \textbf{275}, 424
(2000); R. K. Jackson and M. I. Weinstein, J. Stat. Phys. \textbf{116}, 881
(2004); V. S. Shchesnovich, B. A. Malomed, and R. A. Kraenkel, Physica D
\textbf{188}, 213 (2004); D. Ananikian and T. Bergeman, Phys. Rev. A \textbf{%
73}, 013604 (2006); C. Wang, P. G. Kevrekidis, N. Whitaker and B. A.
Malomed, Physica D \textbf{327}, 2922 (2008); E. W. Kirr, P. G. Kevrekidis,
E. Shlizerman, and M. I. Weinstein, SIAM J. Math. Anal. 40, 566 (2008).

\bibitem{Fatkhulla} F. Kh. Abdullaev, A. M. Kamchatnov, V. V. Konotop, and
V. A. Brazhnyi, Phys. Rev. Lett. \textbf{90}, 230402 (2003); D. E.
Pelinovsky, P. G. Kevrekidis, and D. J. Frantzeskakis, Phys. Rev. Lett.
\textbf{91}, 240201 (2003); F. Kh. Abdullaev and J. Garnier, Phys. Rev. A
\textbf{70}, 053604 (2004); D. E. Pelinovsky, P. G. Kevrekidis, D. J.
Frantzeskakis, and V. Zharnitsky, Phys. Rev. E \textbf{70}, 047604 (2004).

\bibitem{RMP} Yu. S. Kivshar and B. A. Malomed, Rev. Mod. Phys. \textbf{61},
763 (1989).

\bibitem{PTIST} P. G. Kevrekidis, G. Theocharis, D. J. Frantzeskakis and B.
A. Malomed, Phys. Rev. Lett. \textbf{90}, 230401 (2003); L. D. Carr and J.
Brand, \textit{ibid}. \textbf{92}, 040401 (2004); V. P. Barros, M. Brtka, A.
Gammal, and F. Kh. Abdullaev, J. Phys. B: At. Mol. Opt. Phys. \textbf{38},
4111 (2005).

\bibitem{review} R. Carretero-Gonz\'{a}lez, D. J. Frantzeskakis, and P. G.
Kevrekidis, Nonlinearity \textbf{21}, R139 (2008).

\bibitem{Spain} J. Belmonte-Beitia, V. V. Konotop, V. M. P\'{e}rez-Garc\'{\i}%
a, and V. E. Vekslerchik, Chaos, Solitons \& Fractals \textbf{41}, 1158
(2009).

\bibitem{Carr-rep} L. D. Carr, C. W. Clark, and W. P. Reinhardt, Phys. Rev.
A \textbf{62}, 063610 (2000); J. C. Bronski, L. D. Carr, B. Deconinck, J. N.
Kutz, and K. Promislow, Phys. Rev. E \textbf{63}, 036612 (2001).

\bibitem{Carr-attr} L. D. Carr, C. W. Clark, and W. P. Reinhardt, Phys. Rev.
A \textbf{62}, 063611 (2000); J. C. Bronski, L. D. Carr, R. Carretero-Gonz%
\'{a}lez, B. Deconinck, J. N. Kutz, and K. Promislow, Phys. Rev. E \textbf{64%
}, 056615 (2001).

\bibitem{Seaman} B.~T.~Seaman, L.~D.~Carr, and M.~J.~Holland,
%Period doubling, two-color lattices, and the growth of swallowtails in
%Bose-Einstein condensates.
Phys. Rev. A \textbf{72}, 033602 (2005).

\bibitem{Kamch} G. A. El, A. Gammal, and A. M. Kamchatnov, Phys. Rev. Lett.
\textbf{97}, 180405 (2006).

\bibitem{delta1} D. Witthaut, S. Mossmann, and H. J. Korsch, J. Phys. A;
Math. Gen. \textbf{38}, 1777 (2005).

\bibitem{delta2} T. Mayteevarunyoo, B. A. Malomed, and G. Dong, Phys. Rev. A
\textbf{78}, 053601 (2008).

\bibitem{KA} Y. S. Kivshar and G. P. Agrawal, \textit{Optical Solitons}
(Academic Press: San Diego, 2003).

\bibitem{trapping} L. Fallani, C. Fort, and M. Inguscio, Rivista Nuovo Cim.
\textbf{28}, 1 (2005).

\bibitem{painting} K. Henderson, C. Ryu, C. MacCormick, and M. G.I Boshier,
New J. Phys. \textbf{11}, 043030 (2009).

\bibitem{Flach} S. Flach, Y. Zolotaryuk, and K. Kladko, Phys. Rev. E \textbf{%
59}, 6105 (1999); I. V. Barashenkov, O. F. Oxtoby, and D. E. Pelinovsky,
\textit{ibid}. E \textbf{72}, 035602 (2005); D. E. Pelinovsky, Nonlinearity
\textbf{19}, 2695 (2006); S. V. Dmitriev, P. G. Kevrekidis, A. A.
Sukhorukov, N. Yoshikawa, and S. Takeno, Phys. Lett. A \textbf{356}, 324
(2006).

\bibitem{EfrSivilChrist09} N. K. Efremidis, G. A. Siviloglou, and D. N.
Christodoulides, Phys. Lett. A \textbf{373}, 4073 (2009).

\bibitem{AblSegur} M.~J.~Ablowitz, and H.~Segur, \textit{Solitons and the
Inverse Scattering Transform} (SIAM, Philadelphia, 1981).

\bibitem{Yunakovsky} A.~D.~Yunakovsky, \textit{Simulation of the Nonlinear
Schr\"{o}dinger Equation}. (IPF RAN, Nizhny Novgorod, IPF RAN, 1995) (in
Russian); G.~M.~Fraiman, E.~M.~Sher, A.~D.~Yunakovsky, and W.~Laedke,
%Long-term evolution of strong 2-D NSE turbulence.
Physica D \textbf{87}, 325%--334
(1995); Ya.~L.~Bogomolov and A.~D.~Yunakovsky, \textit{Split-step Fourier
method for nonlinear Schr\"{o}dinger equation}, in: Proceedings of the
International Conference, ``Days on Diffraction", p. 34%--42
(2006) (DOI: 10.1109/DD.2006.348170; see also:
http://www.mende%
\-ley.com/c/81983606/Bogomolov-2006-Splitstep-Fourier-method-for-nonlinear-Schr%
\"{o}dinger-eq\-ua\-tion/)

\bibitem{KondrMiller} I.~G.~Kondrat'ev and M.~A.~Miller, %\textit{Study of
%waves in linear inhomogeneous media using the solutions of certain
%nonlinear equations},
Izv. VUZov, Radiofizika \textbf{IX}, 910 (1966) [in Russian; English
translation: Sov. J. Radiophys. Quantum Electr. \textbf{IX}, 532 (1966)].

\bibitem{Snyder} A. W. Snyder, D. J. Mitchell, L. Poladian, and F.
Ladouceur, Opt. Lett. \textbf{16}, 21 (1991).

\bibitem{Miles81} J.~W.~Miles, Tellus \textbf{33}, 397 (1981);
T.~R.~Marchant, and N.~F.~Smyth, IMA J. Appl. Math. \textbf{56}, 157 (1996);
A.~V.~Slyunyaev, ZhETF \textbf{119}, 606 (2001) [in Russian; English
translation: JETP \textbf{92}(3) 529 (2001)]; R.~Grimshaw, D.~Pelinovsky,
E.~Pelinovsky, and A.~Slyunyaev, Chaos \textbf{12} 1070 (2002).

\bibitem{Apel07} L.~A.~Ostrovsky, and Y.~A.~Stepanyants, Chaos, \textbf{15},
037111~(2005). J.~Apel, L.~A.~Ostrovsky, Y.~A.~Stepanyants, and J.~F.~Lynch,
J. Acoust. Soc. Am. \textbf{121}, 695 (2007).

\bibitem{Pelinovsky-Grimshaw} D.~E.~Pelinovsky, and R.~H.~J.~Grimshaw, Phys.
Let.~A. \textbf{229}, 165 (1997); R.~Grimshaw, A.~Slunyaev, and
E.~Pelinovsky, Chaos, (2010) -- to be published.

\bibitem{SlyunPel99} A.~V.~Slyunyaev and E.~N.~Pelinovskii, %Dynamics of
%large-amplitude solitons.
ZhETF \textbf{116}, 318 (1999) [in Russian; English translation: JETP
\textbf{89}(1) 173 (1999)].

\bibitem{WangHao09} Y.~Wang, and R.~Hao,
%Exact spatial soliton solution for
%nonlinear Schr\"{o}dinger equation with a type of transverse nonperiodic
%modulation.
Opt. Commun. \textbf{282} 3995 (2009).

\bibitem{dark} G. Theocharis, D. J. Frantzeskakis, P. G. Kevrekidis, B. A.
Malomed, and Y. S. Kivshar, Phys. Rev. Lett. \textbf{90}, 120403 (2003).
\end{thebibliography}
\end{document}